\documentclass[sigconf]{acmart}
\usepackage{CJKutf8}
\usepackage{verbatim}
\usepackage{spverbatim}
\usepackage{fancyvrb}
\usepackage{fvextra}
\AtBeginDocument{%
  \providecommand\BibTeX{{%
    \normalfont B\kern-0.5em{\scshape i\kern-0.25em b}\kern-0.8em\TeX}}}

\setcopyright{acmcopyright}
\copyrightyear{2018}
\acmYear{2018}
\acmDOI{XXXXXXX.XXXXXXX}

\acmConference[Conference acronym 'XX]{Make sure to enter the correct
  conference title from your rights confirmation emai}{June 03--05,
  2018}{Woodstock, NY}
\acmPrice{15.00}
\acmISBN{978-1-4503-XXXX-X/18/06}

\begin{document}
\title{AI-Gadget Kit: Integrating Swarm User Interfaces with LLM-driven Agents for Rich Tabletop Game Applications}

%  \author{Yijie Guo}
%  \authornote{Both authors contributed equally to this research.}
%  \email{trovato@corporation.com}
%  \orcid{1234-5678-9012}
%  \authornotemark[1]
%  \email{webmaster@marysville-ohio.com}
%  \affiliation{%
\author{Yijie Guo}
 \authornote{Equal contribution.}
\affiliation{%
  \institution{Tsinghua University}
  \city{Beijing}
  \country{China}}
   \authornotemark[0]
  \email{guoyijie.sh@gmail.com}

  \author{Zhenhan Huang}
    \authornotemark[1]
\affiliation{%
  \institution{University of Tsukuba}
  \city{Tsukuba}
  \country{Japan} 
}\email{zhenhan.email.jp@gmail.com}

\author{Ruhan Wang}
  \authornotemark[1]
\affiliation{%
  \institution{Tsinghua University}
  \city{Beijing}
  \country{China}
}\email{wangrh22@mails.tsinghua.edu.cn}

\author{Zhihao Yao}
\affiliation{%
 \institution{Tsinghua University}
  \streetaddress{30 Shuangqing Rd}
  \city{Beijing}
  \country{China}}
  \email{yaozh_h@outlook.com}

\author{Tianyu Yu}
\affiliation{%
  \institution{Tsinghua University}
  \city{Beijing}
  \country{China}}
  \email{yty21@mails.tsinghua.edu.cn}

\author{Zhiling Xu}
\affiliation{%
  \institution{Tsinghua University}
  \city{Beijing}
  \country{China}}
  \email{xzl23@mails.tsinghua.edu.cn}

\author{Xinyu Zhao}
\affiliation{%
  \institution{Tsinghua University}
  \city{Beijing}
  \country{China}}
  \email{xyzhao23@mails.tsinghua.edu.cn}

\author{Xueqing Li}
\affiliation{%
  \institution{Tsinghua University}
  \city{Beijing}
  \country{China}}
  \email{li-xq23@mails.tsinghua.edu.cn}

\author{Haipeng Mi}
\affiliation{%
  \institution{Tsinghua University}
  \city{Beijing}
  \country{China}}
  \email{mhp@tsinghua.edu.cn}

\renewcommand{\shortauthors}{Trovato and Tobin, et al.}

\begin{abstract}
While Swarm User Interfaces (SUIs) have succeeded in enriching tangible interaction experiences, their limitations in autonomous action planning have hindered the potential for personalized and dynamic interaction generation in tabletop games. Based on the AI-Gadget Kit we developed, this paper explores how to integrate LLM-driven agents within tabletop games to enable SUIs to execute complex interaction tasks. After defining the design space of this kit, we elucidate the method for designing agents that can extend the meta-actions of SUIs to complex motion planning. Furthermore, we introduce an add-on prompt method that simplifies the design process for four interaction behaviors and four interaction relationships in tabletop games. Lastly, we present several application scenarios that illustrate the potential of AI-Gadget Kit to construct personalized interaction in SUI tabletop games. We expect to use our work as a case study to inspire research on multi-agent-driven SUI for other scenarios with complex interaction tasks.
\end{abstract}





\begin{teaserfigure}
  \includegraphics[width=\textwidth]{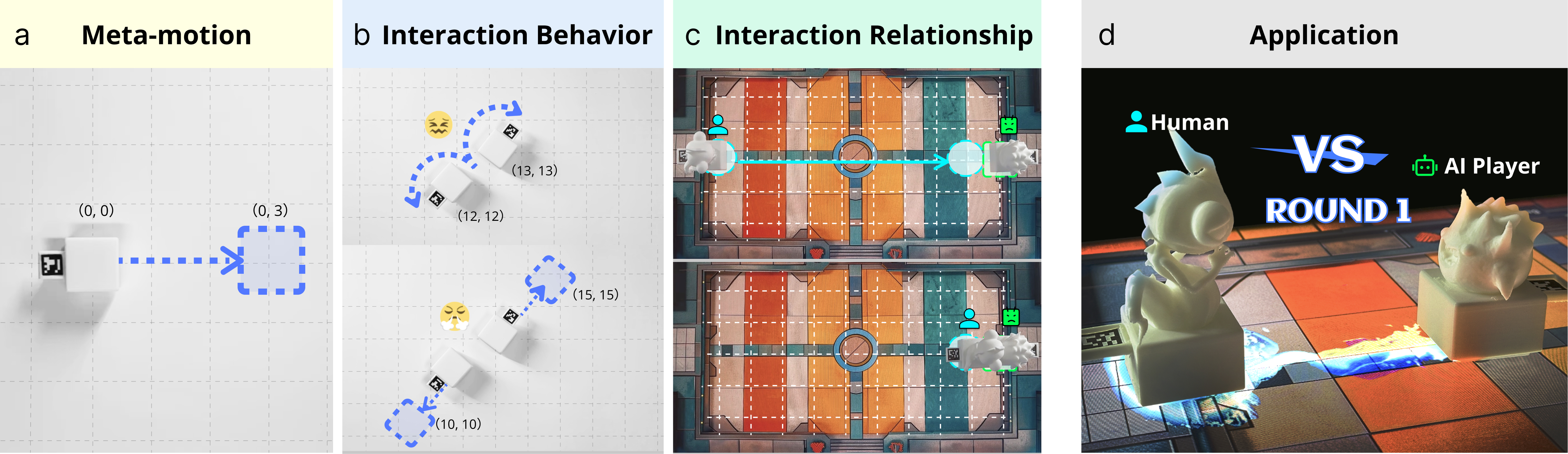}
  \caption{Building a tabletop game with a multi-agent system enabled by AI-Gaget Kit. a) Extend the meta actions of SUIs to gadgets' complex motion planning, b) interaction behavior generation, c) interaction relationship management. d) Robotic gadget plays a turn-based strategy game with a human player.}
  \Description{Building a tabletop game with a multi-agent system enabled by AI-Gaget Kit. a) Extend the meta actions of SUIs to gadgets' complex motion planning, b) interaction behavior generation, c) interaction relationship management. d) Robotic gadget plays a turn-based strategy game with a human player.}
  \label{fig:teaser}
\end{teaserfigure}

\received{20 February 2007}
\received[revised]{12 March 2009}
\received[accepted]{5 June 2009}

\maketitle

\section{Introduction}
%motion抽象的复杂运动
%behaviors具体的复杂运动
%action抽象的一次运动
%movement具体的运动

% 集群机器人界面（SUI）作为一种新兴的人机交互界面，已经吸引了大量研究者的关注。
% 其通过集群的运动协作、驱动实体物体运动、或驱动自身形变等方式，在多种场景展现出了潜在的应用价值，例如信息实体可视化[shapebot]、远程协作[holo]、教学[sketchReality]等。
% 然而，现有研究中的集群机器人界面在应用时，大多局限于预先编程好的一组交互任务。即用户可以通过不断编写新的机器人行动规划，以应对新的交互任务，但这种方式难以动态地应对实际使用时复杂多变的、预先编程范围之外的、新的交互任务。

Swarm User Interface (SUI) as an emerging interface has garnered significant attention from researchers. 
Researchers have explored multiple application scenarios of SUI, such as data physicalization\cite{shapebots}, remote collaboration\cite{holobots}, and educational purposes\cite{sketched-reality}, based on the unique tangible interaction behaviors of SUI, such as collaborative motion, objects actuation, or self-shape changes.
However, most existing SUIs utilize a pre-programmed set of action planning rules to execute different interaction tasks during use, i.e., users need to program the action each time they face a new interaction task.
This approach struggles to adapt to real-world tasks that are usually more complex, dynamic, and possibly changeable beyond the pre-programmed scope.

% 近年来，大语言模型（LLM）技术为机器人的控制与规划带来了新的机遇。
% 大语言模型能够基于自身的知识库，对复杂的情境、任务和语义进行深度理解和推理，从而动态地生成回答或解决方案。
% 以往的机器人行动规划通常依赖单一决策模型，即预测机器人在交互中的每一步移动和行动编排（to make every step movement prediction）[audio-visual-embodied]、[Enviroment-Agnostic]。
% 近年有研究使用大语言模型，特别是LLM驱动的智能体技术，驱动机器人面对更广泛的交互任务进行行动规划[navgpt][discussnav]。例如，DiscussNav[9]通过使用多个具有不同特长的LLM/VLM智体，轮流对机器人在复杂场景下的导航进行决策。
% 然而，这些工作的研究对象通常是单一机器人，多机器人，例如集群机器人界面，利用大语言模型面对复杂交互任务进行协调的行动规划的研究依然较少。

In recent years, Large Language Models (LLM) have shown benefits for robotic motion control and planning. 
Based on the embedded knowledge of LLMs, they are capable of understanding and reasoning about the complex contexts in different tasks, thus dynamically generating responses, e.g., motion planning, based on the contexts in the tasks.
Traditional methods for robotic motion planning included using one-decision models, which relied on the prediction of every step of the robot's movement\cite{audio-visual-embodied,Environment-agnostic}.
Recent research has explored using LLMs, particularly LLM-driven agents, to conduct robotic motion planning for more complex interaction tasks.
For instance, DiscussNav\cite{discussnav} has used multiple LLM agents with different expertise to make decisions for robotic navigation in a complex interior scenario.
Nevertheless, most of these works focused on single-robot systems.
Research on multiple-robot systems, e.g., SUI, and how to use LLMs to assist the motion planning of these systems for complex interaction tasks still remains unexplored.

% 在集群机器人的众多应用场景中，桌面游戏是一个典型的具有复杂交互任务的场景。
% 已有研究探索实体交互界面、集群机器人界面探索在桌面游戏中的应用，旨在丰富桌面游戏的互动性、趣味性，例如通过使用机器人在游戏中与玩家进行各种实体交互，实现包括机器人玩家（19）（15）、机器人Game Master（25）、或自动化的道具（27）（22）的新颖功能。
% 然而，这些研究中的机器人行动规划依然基于预先的编程，尤其难以应对桌游中广泛存在的复杂交互任务，例如对游戏复杂剧情、玩家不确定性的操作、用户即兴或情绪化的表达等的理解和反应。

Among the various applications of SUI, the tabletop game is a typical scenario that contains versatile complex interaction tasks. 
Existing research has explored the application of tangible and swarm user interfaces in tabletop games to enhance interactivity and enjoyment, such as using robots to facilitate embodied AI players\cite{gambit,icat}, robotic game masters \cite{social-mediator}, and automated gadgets\cite{remote-you,practical}.
However, the action planning for robots in these studies still relies on pre-programmed rules, which makes it challenging to execute the complex interaction tasks in tabletop games, such as understanding and reacting to complex game narratives, improvised decisions of players, or emotional expressions from players.

% 基于以上gap，我们希望以集群机器人的桌面游戏场景为案例，探索大语言模型在集群机器人面对复杂交互任务进行行动规划时的应用。
% 在本文中，我们提出AI-gadget Kit，一个多智能体驱动的集群机器人桌面游戏系统，以实现用户在桌面游戏中的动态的、复杂的交互任务。
% 我们首先介绍了AI-gadget Kit的系统结构，包括一组用于承载桌面游戏实体交互的现有集群机器人平台，和用于执行游戏、并生成集群机器人行动规划的多智能体系统。
% 我们其次介绍了多智能体系统的设计，包括一系列单一机器人的基础运动元素、两个用于复杂行动规划的智能体、以及强化桌游场景理解、反应能力的一系列附加提示词，包括关于桌游中典型交互行为与交互关系的提示词。
% 最后，我们使用AI-gadget Kit演示了4种交互场景，用于展示多智能体系统驱动的集群机器人，在完成桌面游戏复杂交互任务时的效果。
% 我们希望通过本文启发 多智能体系统驱动的集群机器人 在 其他具有复杂交互任务的场景中的 研究与应用。

%大语言模型在集群机器人面对复杂交互任务进行行动规划时的应用。
% 多智能体系统驱动的集群机器人完成桌面游戏复杂交互任务
% 智能体驱动的集群机器人作为桌游中的gadget实现复杂交互的效果。

In this paper, we aimed to use the tabletop game as a case study to explore the application of LLMs on action planning of SUI in scenarios with complex interaction tasks.
We proposed AI-gadget Kit, a multi-agent SUI tabletop gaming system, which is designed to facilitate dynamic and complex interaction tasks in tabletop games.
We first introduced the system architecture of the AI-gadget Kit, which includes a set of swarm robots based on existing platforms to perform the gadget behaviors, and a multi-agent system responsible for executing the game and generating action plans for the swarm robots.
We then elaborated the design of the multi-agent system, comprising a series of meta-motions for individual robots, two LLM-based agents for complex action planning, and a set of add-on prompts aimed at reinforcing the understanding and reacting capabilities of the agents.
At last, we demonstrate four application examples using AI-gadget Kit to showcase the effect of the multi-agent-driven SUI on executing complex interaction tasks in tabletop games.
Through this work, we aim to inspire the research of multi-agent-driven SUI on other scenarios with complex interaction tasks.

In summary, the contribution of this paper includes:
\begin{enumerate}
    \item AI-gadget Kit, a multi-agent SUI tabletop gaming system, which consists of a series of meta-motions for individual robots, two LLM-based agents for complex action planning, and a series of add-on prompts tailored to the tabletop gaming scenario to enhance the understanding capability of the multi-agent system. 
    \item A set of application examples using AI-gadget Kit on tabletop games, which demonstrates the effect of agent-driven swarm robots as gadgets in tabletop games for complex interaction tasks.
\end{enumerate}

\section{Related work}

\subsection{Swarm User Interface}

% 与传统TUI相比[ken]，引入物理移动机器人作为驱动实体的SUI[phycial objects]，在交互中提供了灵活、扩展的物理交互空间，以及多感官的交互体验。
% HCI的研究人员此前已经探索了多种使用SUI作为用户界面的可能性，例如，SwarmHaptic（8）使用在平面上移动的小型轮式集群机器人构建了一种新型触觉界面，Rovables（11）可在穿着的衣物上进行自主移动，为移动体感设备提出了一个涵盖传感、驱动和接口的交互空间，ShapeBots（10）使可自我变形的机器人群单独或集体改变其配置，以在物理空间中展示信息，Ractile（9）探索了一种利用直接物理操作对集群用户界面进行编程的新方法。Holobots提出了a mixed reality remote collaboration system that augments holographic telepresence with synchronized mobile robots[holo]，HERMITS introduce a modular interaction architecture for swarm user interface.

% 此外，SUI在教育与娱乐场景中也已经展示了广泛的应用前景，Sony使用可编程的小型机器人Toio（12）来让小学到成年的学习者进行逻辑思考，了解日常生活和周围社会问题的解决机制并将其付诸实践。Thymo机器人为各年龄段的学习者提供全面、吸引人的 STEM 教育（13）. 

% 利用这些界面在交互形态和控制的统一性，以及丰富交互性，我们可以驱动桌面游戏中的多种不同gadget，为它们赋予个性化的交互，从而增强交互过程中的体验。然而，当前SUI面临的一大挑战是他们主要支持既定编程的情景。在桌面游戏中，这种既定编程很难在不同情境下支持复杂多样的用户指令与需求。

Compared to traditional Tangible User Interfaces (TUI), Swarm User Interfaces (SUI) introduce multiple moving robots that enable collaborative motion, providing a flexible and extensive physical interaction space and multi-modal interaction experiences. 

Researchers in Human-Computer Interaction (HCI) have explored various applications of SUI.
For instance, SwarmHaptic\cite{SwarmHaptics} utilized small wheeled swarm robots moving on a flat surface to construct a novel tactile interface.
Rovables\cite{Rovables} provided a series of robots that were capable of autonomous movement on wearable clothing, proposing an interaction space for sensing and actuation on wearables.
ShapeBots\cite{shapebots} enabled a group of self-deforming robots to individually or collectively change their configuration to display information in physical space. 
Additionally, Holobots\cite{holobots} proposed a mixed-reality remote collaboration system augmenting holographic telepresence with synchronized mobile robots.
Out of academia, SUIs have also demonstrated extensive application prospects in education and entertainment scenarios.
Sony employs programmable small robots called Toio\footnote{https://toio.io/} to engage learners from elementary to adult in logical thinking and learning programming
Thymo\footnote{https://www.thymio.org/} robots provide STEM education for learners of all ages.

However, most existing SUIs utilize a pre-programmed set of action planning rules to execute different interaction tasks during use.
For example, although Holobots creatively proposed six interaction types for remote collaboration, constrained by predetermined programming, they struggled to dynamically generate personalized tactile feedback based on users' flexible needs during actual usage. 
Thus, a system that is capable of understanding and reacting to complex interaction tasks will significantly improve the generalizability of interaction behaviors of SUIs in these works.
% althogh
% 在此基础上，当前SUI面临的一大挑战是对情景的泛化能力。
% 例如，虽然holobots创造性地针对有形远程协作提出了6类交互，但受限于既定编程，他们很难在实际使用过程中依据用户不同的需求动态地生成个性化的触觉反馈。为SUI提供一套能够理解并规划这些交互行为的系统，将会帮助相关的工作提供更丰富的交互性，和泛用性。

% The Swarm User Interface (SUI) as an emerging interface has garnered significant attention from researchers. 
% Researchers have explored multiple application scenarios of SUI, such as data physicalization[shapebot], remote collaboration[holo], and educational purposes[sketchReality], based on the unique tangible interaction behaviors of SUI, such as collaborative motion, objects actuation, or self-shape changes.
% However, most existing SUIs utilize a pre-programmed set of action planning rules to execute different interaction tasks during use, i.e., users need to program the action each time they face a new interaction task.
% This approach struggles to adapt to real-world tasks that are usually more complex, dynamic, and possibly changeable beyond the pre-programmed scope.

% However, a major challenge currently faced by SUIs is their ability to generalize across scenarios. 

\subsection{Agents for Action Planning}
% 在机器人控制中，研究者希望向机器人智能体发布高级指令，智能体可以自动将高级指令转化为低级动作并由机器人执行，而不需要人类直接对机器人的低级动作进行编辑。例如，Skill Transformer[1]通过一个基于Transformer架构的神经网络模型对机器人的低级动作进行预测，使机器人能够在复杂环境中根据指令给定的目标物体和目标位置，完成移动物体的具身任务。大语言模型（LLM）出现后，研究者试图利用LLM强大的自然语言理解能力，处理泛化、自然语言的具身指令。March in Chat[5]通过agent与LLM、VLM进行交互，根据模糊自然语言指令在日常活动场景中进行导航。VoxPoser[6]通过LLM对场景中各物体可能对完成指令产生的效益和损失进行估计，生成场景的3D值图，进而得到机器人的运动轨迹。
In the domain of robotics, researchers aim to issue high-level instructions to robotic agents. These agents automatically translate the instructions into low-level actions for execution by robots, eliminating the need for humans to manually program. The Skill Transformer\cite{skilltrasnformer}, leveraging a neural network model based on the Transformer architecture\cite{Attention}, predicts low-level actions for robots, enabling them to accomplish embodied tasks of moving objects to specified targets and locations in complex environments. With the advent of Large Language Models (LLMs), researchers have sought to harness LLMs' robust natural language understanding capabilities to process generalized, natural language-based embodied instructions. For example, March in Chat\cite{March_in_chat} interacts with agents, LLMs, and VLMs to navigate daily activity scenes based on vague natural language instructions. VoxPoser\cite{voxposer} estimates the potential benefits and losses of objects in a scene towards fulfilling an instruction using LLMs, generating a 3D value map of the scene to derive the robot's trajectory.

% 研究者同样注意到，相对于单一智能体独自进行决策，多个智能体协同决策一个机器人的行动，可以使机器人适应更加复杂的场景。DiscussNav[9]被用于解决机器人在复杂场景下的导航问题。机器人进行每一步行动时，多个具有不同特长的LLM/VLM智体轮流参与决策，提高了机器人导航的泛化能力。
Researchers have also recognized that collaborative decision-making among multiple agents for a robot's actions can enable adaptation to more complex scenarios compared to decisions made by a single agent alone. DiscussNav\cite{discussnav} is used to address navigation problems in complex scenes, where robots take each step with the involvement of multiple LLM/VLM agents with different specialties, enhancing the robots' generalization ability in navigation.

%尽管现有研究在具身智能体的行动规划和硬件控制上取得了较好的表现。然而，现有研究他们对具身任务规划的验证基本圈定于日常情境下的操作任务[1, 2, 3, 4, 6]或导航任务[5, 6, 9]，而在特定任务场景、虚构叙事场景等特殊场景下的验证应用较为缺少。此外，它对于多机器人驱动的用户界面在交互上的规划也有待验证。探索LLM-based agent与SUI的结合将会为相关工作拓宽机器人的应用场景与交互案例。
Despite existing research has shown feasible performance in action planning and robotic control for embodied agents, their validation of task planning for embodied tasks has been primarily confined to operational tasks in daily routine scenes \cite{skilltrasnformer,voxposer} and navigation tasks\cite{discussnav} with less emphasis on other certain genres of task scenarios, e.g., fictional narrative settings. Moreover, its interaction planning capabilities in user interfaces driven by multiple robots also require validation. Exploring the integration of LLM-based agents with SUI will broaden the application scenarios and interaction cases for related work in robotics.

\subsection{Robots in Tabletop Game}
In recent years, researchers have explored the application of tangible and swarm user interfaces in tabletop games to enhance interactivity and enjoyment.

With the emerging trend of robots being small and versatile, more and more robots are used as embodied AI players or gadgets in tabletop games.
For example, Brock et al.\cite{remote-you} utilized the robot Haru to simulate the behavior of remote human players. 
Researchers also used robots to serve as robotic gadgets in the game, such as creating chess that can move automatically to enable novel and compelling interaction experiences\cite{practical}.
Sparkybot\cite{sparkybot} allowed children to use mobile robots as different actors in storytelling games to enhance children's creativity.

These works, that use SUI for tabletop games, provide rich user interaction spaces. 
However, the action planning for robots in these studies still relies on pre-programmed rules, which makes it challenging to execute the complex interaction tasks in tabletop games, such as understanding and reacting to complex game narratives, improvised decisions of players, or emotional expressions from players.
In this work, we aimed to leverage LLM-based agents to assist action planning for SUI, in order to execute complex interaction behaviors of the gadgets in tabletop games.

% 随着桌面机器人种类的丰富和小型化趋势，越来越多的机器人被用在桌面游戏中的道具或执行者。例如，Brock.et.al使用了机器人Haru来模拟和反映远程人类玩家的行为（27）。
% 此外，机器也可以作为新型游戏道具，例如每个棋子具有自动移动机制的新型可交互棋盘（22）被构建，以增强棋类游戏对玩家的吸引力。
% Sparkybot 允许儿童自定义移动机器人机器人在讲故事中的角色和交互行为以增强儿童创造力[sparkybot]. 
% AutomaTiles （21）提出了一种小型桌面集群机器人，作为新型的游戏平台，它将计算智能融入桌面游戏中。

% 这些将桌游与机器人或群体用户界面（SUI）结合的案例，提供了丰富的用户交互空间。
% 然而，这些研究中的机器人行动规划依然基于预先的编程，尤其难以应对桌游中广泛存在的复杂交互任务，例如对游戏复杂剧情、玩家不确定性的操作、用户即兴或情绪化的表达等的理解和反应。
% 本文旨在深入探讨如何借助基于大型语言模型（LLM）的智能体辅助SUI进行行动规划。
% 基于此，我们可以驱动桌面游戏中的多种不同gadget，为它们赋予个性化的交互。

\section{System Architecture of AI-Gaget Kit}

\begin{figure}[!htbp]
    \centering
    \includegraphics[width=\linewidth]{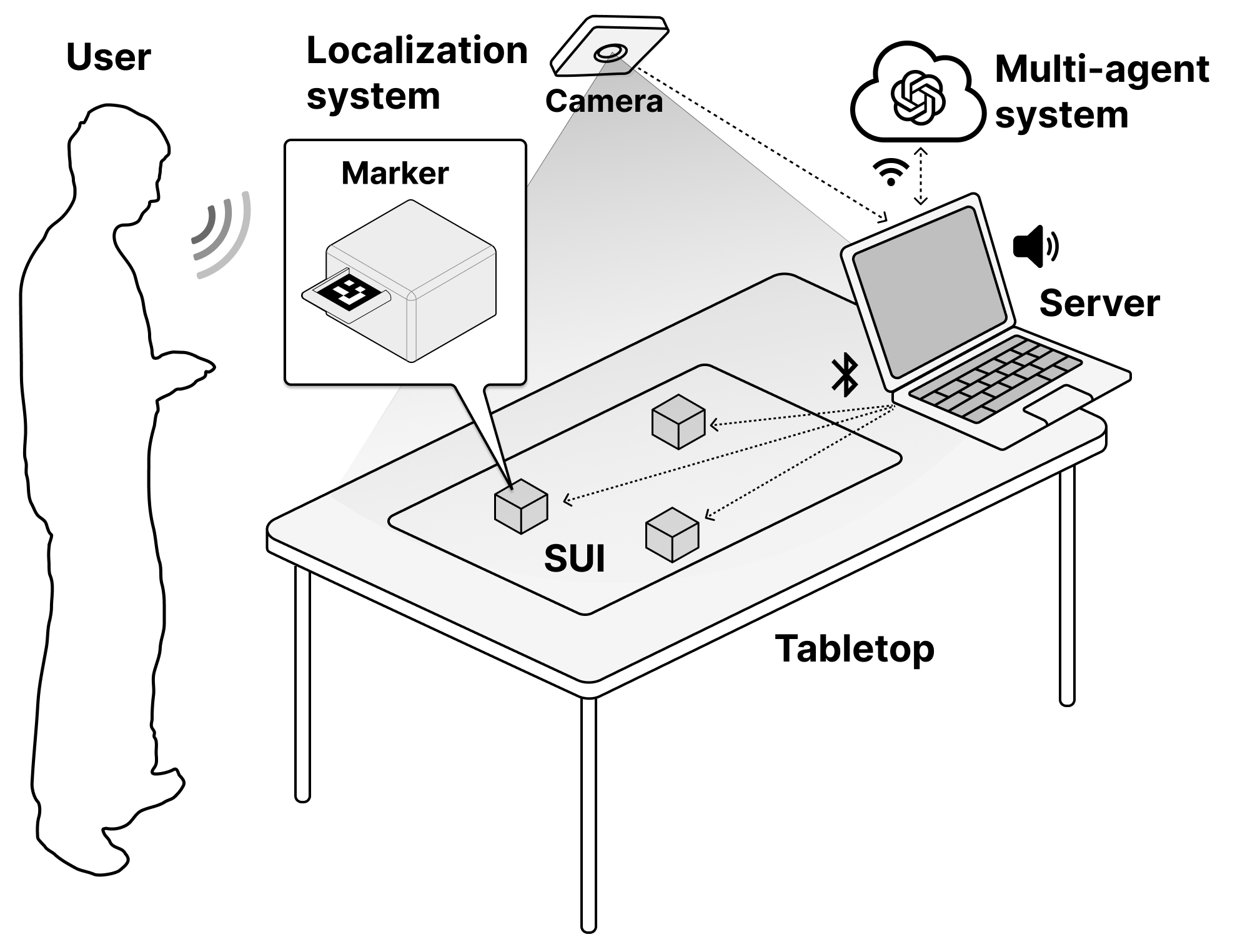}
    \caption{System Architecture of AI-Gadget Kit.}
    \label{fig:systemarchitecture}
\end{figure}

% AI-Gadget Kit系统如图？所示，包括一组SUI平台、定位系统、服务器、以及基于LLM的一套多智能体系统。
% 其中SUI平台包含多个可独立运动的机器人，用于驱动桌面游戏中的gadget的各种行为。
% 定位系统包括一台摄像机和位于机器人上的定位标识，用于在使用时获取每一台机器人的位置和朝向数据。
% 服务器一方面获取用户的文本或语音输入命令、以及来自定位系统的机器人方位数据，发送至LLM多智能体系统；另一方面也接收LLM所返回的驱动信息，经处理后向机器人发送运动指令，并播放音效等。
% LLM的多智能体系统负责游戏核心的交互计算，基于游戏的规则和知识，对用户的输入命令进行反应，并推理游戏中gadget的行为，生成用于控制SUI机器人的行动序列。
The system architecture of AI-Gadget Kit, shown in Figure\ref{fig:systemarchitecture}, consists of an SUI platform, a localization system, a server, and a LLM-based multi-agent system. 
The SUI platform includes multiple independent robots, which are used to actuate various gadget behaviors in tabletop games. 
The localization system comprises a camera and multiple on-robot markers, used to obtain the position and orientation of each robot. 
The server obtains users' text or verbal input commands, as well as robots' position and orientation data, and then sends them to the LLM-based multi-agent system for information processing. 
The server also receives the actuations generated by the multi-agent system, playing sound effects, or sending action sequences to the robots in the SUI platform.
The LLM-based multi-agent system is responsible for the execution of the core game interactions. 
Based on the game's rules and knowledge, the multi-agent system responds to user commands, reasons the gadget behaviors in the game, and then generates the action sequences to control the SUI robots.

% 具体地，在本次工作中，我们选用一张1m*1m的桌面被用作为SUI进行交互的场地，经过预先的校准和设置，该活动区域被划分为30*30的世界坐标系，分别以东、以北为x、y轴的正方向。
% 我们选用了Sony的Toio作为SUI中的机器人，其单个机器人的尺寸为3.2cm*3.2cm*2.5cm。
% 我们使用？作为服务器，服务器通过蓝牙向Toio机器人发送数据，通过wifi与LLM进行通信。
% 我们在定位系统中，使用imx415网络IP摄像头，安装在桌面正上方1m处，垂直向下拍摄整个桌面的范围和机器人上的定位码。我们在每一台机器人上粘贴Aruco码[shapebots][hermits][小行星]作为定位码。定位系统工作时，服务器通过RTSP协议获取摄像头拍摄的视频流，并基于Python的Opencv库中的Aruco模块，以30fps追踪每一个机器人的位置和方向。
% 我们基于Gpt4开发了多智能体交互系统。工作时，服务器将用户输入的文字或语音命令转换为文本，并与机器人的方位数据一起上传到云端（AWS），以供agent进行分析处理。多智能体系统的设计见后文。

In this project, we utilized a 1m*1m tabletop as the interaction space. 
The space was divided into a 30*30 coordinate system, with the east and north directions serving as the positive directions of x and y axes, respectively.
We used Sony's Toio robots as the SUI platform. Each individual robot has dimensions of 3.2*3.2*2.5cm.
We used a PC as the server, which communicates with Toio robots via Bluetooth and communicates with the LLM via WiFi.
In our localization system, we employed an imx415 network IP camera, positioned 1m above the tabletop, and we used ArUco codes as the localization markers on the robots. 
The server retrieves the video stream from the camera using the RTSP protocol and uses the Python OpenCV to track the position and orientation of each robot.
We developed the multi-agent system based on GPT4 LLM. The design of the multi-agent system is introduced as follows.

\section{Multi-agent system}
% 在AI-Gaget Kit中，我们使用多智能体系统来负责游戏核心的交互计算。
% 其基于游戏的规则和知识，对用户的输入命令进行反应，并推理游戏中gadget的行为，生成用于控制SUI机器人的行动序列。
% 具体地，其包含一组机器人的meta-action，一组……、一组……，以及由coodinator和controller构成的多智能体系统。
% 具体介绍如下。
In the AI-Gadget Kit, we utilized a multi-agent system to compute the core interaction of the game.
Based on the rules, and knowledge of the game, the system responds to the user's command, reasons the gadget behaviors within the game, and then generates the action sequences to control the SUI robots.

To build the multi-agent system, we first defined a set of meta-actions for each single robot, and then designed a two-agent system, including a \textit{Coordinator} agent and a \textit{Controller} agent, to learn and use those meta-actions for complex motion planning.
We also designed a set of add-on prompts, including prompts for \textit{Interaction behavior planning} and \textit{Interaction relationship planning}, to enhance the agents to understand and react to complex interactions during the game.

\begin{figure*}[!htbp]
    \centering
    \includegraphics[width=\linewidth]{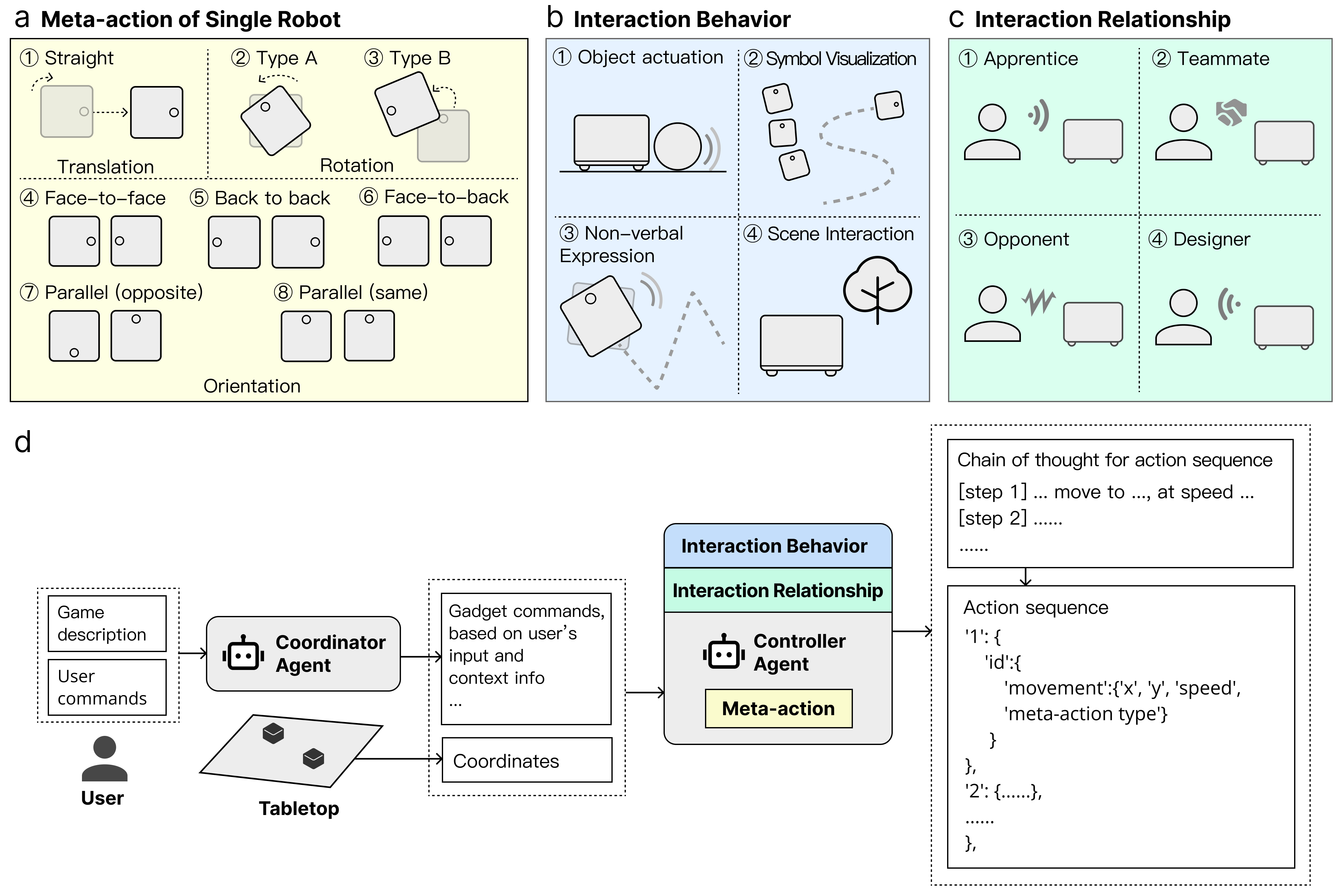}
    \caption{Design space for SUI action planning in tabletop game scenarios through our kit. The kit includes a two-agent system(d) that can generate action sequences for gadgets to complete interaction tasks based on eight meta-actions (a), four types of interaction behaviors (b), and four types of interaction relationships (c).}
    \label{fig:design space}
\end{figure*}

\subsection{Meta-action for Individual Robot}
% 如果想使用Agent为SUI生成其中机器人的behavior plan(行为规划)，我们首先需要明确单个机器人个体所具备的premitive patterns for movement, i.e., meta-action(基本运动方式)。
% Toio机器人依靠连接到左右轮的两台电机实现移动，因此，我们可以通过控制两台电机的转向（正、负）、转速（慢、中、快）、和运动时长（x秒）来实现对机器人movement(行动)的控制。
To leverage the LLM-based agents for planning and generating complex actions for our SUI, we first defined the primitive movement patterns of an individual robot, i.e., meta-actions (Figure \ref{fig:design space}a). 
An individual Toio robot moves by controlling the motors of the two wheels.
Thus, in this work, we controlled the robot's meta-action by controlling the rotation direction (clockwise by default or anti-clockwise), speed (three levels - 10, 20, or 30\footnote{Speed values in Toio platform}), and the duration (x seconds) of each motor.

% 我们将每台机器人的meta-action分为translation(平移运动)和rotation(旋转运动)两种运动moveme，其具体定义如下（如图？）：
% - Rotation:
%   - Rotation A: 机器人两个轮子以相反的方向同速转动一定时长，实现围绕自身中心的原地旋转一定角度。
%   - Rotation B: 机器人一个轮子旋转一定时长、另一个轮子不转，实现围绕自身一侧的旋转一定角度。
% - Translation: 机器人首先通过rotation A将自身角度调整至平移的朝向，再通过两轮的同速、同向转动一定时长，实现直线平移运动一定距离。
% 在机器人执行某一次运动时，服务器将根据机器人预期移动的距离、角度，以及自身轮子、边长尺寸等数据，计算每一次运动所需的时长。
We categorized the robot's meta-actions into two types of movements: \textit{translation} and \textit{rotation}, defined as follows:
\begin{itemize}
    \item \textbf{Rotation:}
    \begin{itemize}
        \item \textbf{Rotation A:} The robot rotates around its center, by spinning its wheels in opposite directions at the same speed for a specified seconds, achieving a precise angle of rotation.
        \item \textbf{Rotation B:} The robot rotates around one side of itself, by spinning only one wheel for a specified seconds, achieving a specific angle of rotation.
    \end{itemize}
    \item \textbf{Translation:} The robot adjusts its orientation to the desired direction through Rotation A, followed by spinning both wheels in the same direction and speed for a few seconds to achieve linear translation over a specific distance.
\end{itemize}
During each movement, the server determines the duration based on a simple calculation of expected translation or rotation displacement and official kinematic data\footnote{https://toio.github.io/toio-spec/en/docs/hardware\_components}.

% 除此以外，考虑到多个机器人的之间的互动，我们为两台机器人设计了五种调整朝向关系的运动，具体包括：面对面、背对背、面对背、平行、反向平行，定义如图？所示。
% 在执行朝向关系调整的运动时，两个机器人从原有位置和角度，分别通过Rotation A，将自身角度调整为最终的朝向关系，位置保持不变。
Furthermore, considering interactions between multiple robots, we designed another type of meta-action to adjust their relative orientations, including face-to-face, back-to-back, face-to-back, parallel, and counter-parallel, defined and illustrated in Figure \ref{fig:design space}a.
The two robots conducting one of these meta-actions adjust their orientations by executing Rotation A for a time.

% 在使用AI-gadget kit时，系统中的agent将根据游戏中对gadget的复杂操作，多次调用translation、rotation和ortientation这三类meta-action、并带入自定义的参数数值，得到机器人的行动序列（action sequence)，用于实现复杂的集群机器人移动（complex motion planning），驱动游戏中的gadget与用户互动。

By repeatedly calling and combining these three types of meta-actions with custom parameters, the agents in the AI-Gadget kit could generate multiple actions in sequences to facilitate the complex motion planning of these robots in SUI.

\subsection{Complex Motion Planning}
% 为了使得SUI可以在桌面游戏中，自行根据用户命令、游戏规则、情景等复杂信息，为机器人生成基于meta-action的行动序列，扮演游戏中的gadget，我们基于大语言模型GPT4创建了一个多智能体（mutil-agent）系统，用以构建AI-Gadget kit（如图x）。
% 我们通过设计不同的提示词分配给LLM，实现扮演不同角色的智能体，智能体相互协作完成基于SUI的桌面游戏任务。
To facilitate the complex motion planning for SUI to execute the gadget behaviors in tabletop games, we developed an LLM-based two-agent system, that aims to understand and react to the game contexts and then generate the action sequences for the robots (Figure \ref{fig:design space}d).
% By assigning expert prompts to the LLM-based agents in the system, the agents are designated to assume varied roles, enabling collaborative efforts among these agents to accomplish tabletop game tasks utilizing SUI.
% 具体地，我们设计了两个agent：协调员（coordinator agent）和控制员（controller agent）。下面将介绍创建过程：
Specifically, we proposed two agents in the system with expert prompts: \textit{Coordinator} and \textit{Controller}:

\textbf{(1) Coordinator Agent.}
% 在桌面游戏中，游戏操作(operation)通常涉及玩家命令(command)输入、游戏情景信息(context)变化等复杂的信息处理。
% 以国际象棋为例，在“命令皇后移动到a1位置”这一行动中，涉及的命令包括“对皇后棋子的操作”，涉及游戏情景信息则包括棋盘的尺寸、皇后的移动规则、是否能够消除对方棋子等。
The operation in a tabletop game typically involves the user's commands and the processing of context information. 
Taking chess as an example, if the user inputs a command of "move the queen to A1", the execution of this command involves the actual movement of the queen, and processing of context information such as the dimensions of the chessboard, the queen's movement rules, and whether an opponent's piece can be captured.
% 我们设计coordinator智能体，负责接收玩家命令，并结合游戏情景信息，解析游戏过程中对于各个gadget的行为。
Hence, we designed a Coordinator agent to respond by processing the players' commands and the game context information, then reasoning the commands for each gadget within this interaction step.
% 通过在coordinator的提示词中实现对coordinator行为的planning，我们让coordinator扮演游戏中的管理员、协调员、裁判等角色。
In the prompts for Coordinator, we asked the agent to act as administrator, coordinator, and referee in the game.
We added the description of a series of duties to the prompts, such as explaining rules, coordinating actions, and updating game states.
% 我们提示词中输入游戏的基本环境参数（如棋盘大小），并让coordinator对玩家输入的任何游戏操作相关的命令作出反应，如解析对gadget的行为、更新游戏情景信息、判断和澄清游戏规则等。
We also inputted the game's environmental settings to the prompts, including the size of the map and the coordinate system.

% 注意到，为了使Coordinator agent能够集中于游戏操作本身，我们采用了reality-agnostic方法，仅让coordinator对游戏操作进行协调代理，而不涉及处理SUI中机器人的现实信息，如位置、运动等参数。
% 其提示词见？。
Note that, to allow the Coordinator agent to focus on game operations, we employed a "reality-agnostic" approach.
Specifically, the Coordinator only facilitates the execution of the game, without dealing with the physical parameters of the SUI robots, such as their locations or next movements. 
We aim to use this method to enhance the Coordinator's understanding and reaction capabilities through efficient prompting of LLM. 
The prompts of the Coordinator are detailed in Supplementary Material.

\textbf{(2) Controller Agent.}
% 为了使用SUI进行桌面游戏，我们仍需要让SUI中的机器人根据玩家的命令和游戏情景信息进行运动(act)和行动(behave)。
% 为此，我们设计controller智能体，负责扮演游戏中的角色，并控制SUI中对应机器人作为Gadget，实现游戏的实体互动。
% 在controller的提示词中，我们令controller从两处获取信息，包括Coordinator agent所输出的对gadget的行为，以及从系统中定位模块获得的当前各期机器人的物理位置坐标数据。
% 接下来，我们让controller在兼顾游戏操作的逻辑和游戏的用户体验的同时，基于上文中定义的meta-action，为需要行动的机器人生成所需的行动序列(action sequence)。
% 其中，我们令该agent建立了一个思维链[https://arxiv.org/abs/2201.11903]的过程，以确保所生成行动序列具有正确的格式，从而与控制机器人的服务器进行通信。
% Agent 的思考过程如下：（1）输出`行动序列的文字描述`和`机器人当前的位置信息`，（2）根据以上信息，输出用于控制机器人运动的`行动序列Python字典`，其包括：本次行动机器人的标识符、以及按顺序构成的每一次meta-action的参数（如位置/角度终点、速度、运动种类等）。
% 此外，我们采用了in-context learning[https://arxiv.org/abs/2101.06804] and few-shot learning的方法，在提示词中给出了一个能够表现上述思考过程的具体案例，以帮助agent快速学习如何生成规定的行动序列。
% 综上，Controller的提示词见？。
To use SUI in tabletop games, we require the robots to plan their motion according to the gadget behaviors in the game. 
To this end, we proposed a Controller agent in our system, which is responsible for embodying characters and generating the action sequence of each robot that represents these characters.
The Controller agent is designed to gather information from two sources: the gadget commands outputted by the Coordinator agent, along with the physical location data of the robots at the given time.
Next, we asked the Controller to generate the action sequences for the robots using the meta-actions, while simultaneously considering the logical flow of the game and the gameplay experience.

To ensure these action sequences are properly formatted for the robot actuation, we designed a Chain-of-Thought (CoT) prompting \cite{COT} for the Controller.
The CoT prompting of the Controller unfolds as follows: (1) Output a textual description of the action sequence of the robots that will move, along with the current location information of these robots; (2) Generate an action sequence based on the aforementioned textural description in the form of a Python dictionary. This Python dictionary encompasses ID for the robots, as well as details for each subsequent meta-action, including destination locations/angles, speeds, types of movement, etc.
Furthermore, we utilized in-context learning and few-shot learning approaches\cite{good_gpt-3}, which provide specific examples in the prompts to demonstrate the process of the aforementioned CoT prompting, to assist the agent in effectively learning how to generate and plan the motion sequences.

% 在使用时，用户首先向系统输入游戏介绍及规则，以声明本场将要进行的游戏。系统基于大预言模型内置的常识，理解游戏、并对游戏进行初始化。
% 随后，用户不断地向系统输入游戏命令，以进行游戏。系统中的两个agent则根据游戏命令和情景信息，推理游戏中gadget的行为，生成用于控制SUI机器人的行动序列，驱动机器人完成游戏中与用户的互动。
During use, users first input the game's description and rules to the system, in order to declare the game to play.
The system then understands and initiates the game based on the inherent knowledge of LLMs.
Then, users continuously input the game commands to the system to engage with the game.
The two agents in the system analyze these commands and the ongoing contextual information of the game, reasoning the gadget behaviors in the game, and then generating the action sequences to actuate the motion of SUI robots, embodying the interactions of the gadgets with users.

\begin{figure*}[!htbp]
    \centering
    \includegraphics[width=\linewidth]{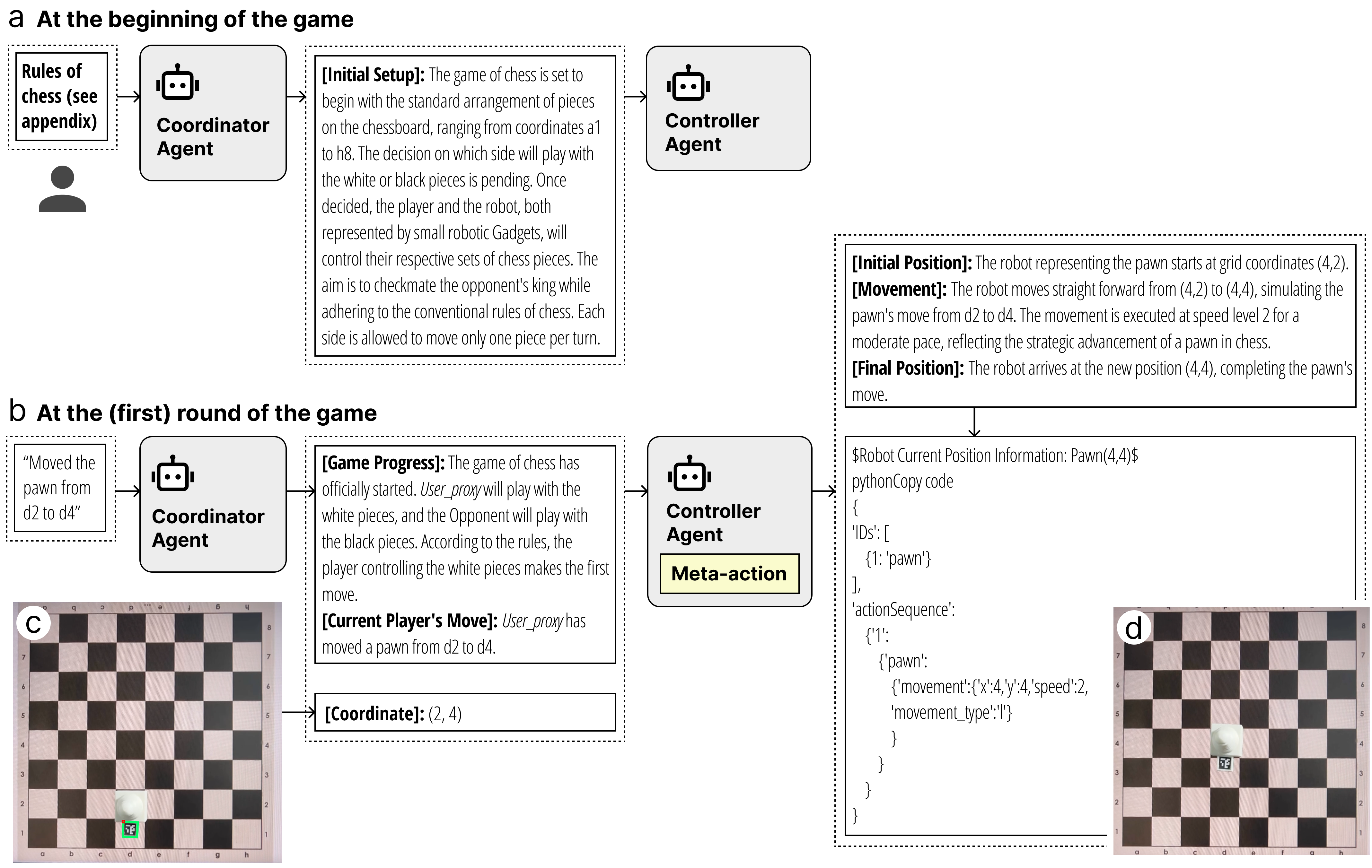}
    \caption{An example using our kit: (a) Users can declare the game they are playing through an introduction and rules entry. (b) Then, they input game commands based on the scenario to interact with the gadget. The gadget generates corresponding action sequences based on a two-agent system within the kit.
}
    \label{fig:design space example}
\end{figure*}
% 下面给出一个具体的测试案例，其中用户希望使用AI-Gadge kit玩国际象棋。
% 首先，用户向系统声明本场游戏为“国际象棋”，当用户输入附录中关于国际象棋的介绍后（见附录），系统中的coordinator便根据用户的输入完成对游戏的初始化设置：

Here we presented a specific case of using the two-agent system to play a chess game. 
The comprehensive process of the interaction was illustrated in Figure \ref{fig:design space example}, including the initiation step of the game and a typical step during one round of the game.
In the initiation step, the user first inputted the game name and a piece of introduction to the system (see Supplementary Material).
The Coordinator then initialized the game based on the user's input and outputted the response as shown in Figure \ref{fig:design space example}a.
Note that in this step, since the user did not input any specific game commands, the Controller agent did not generate specific action sequences, either.
% 接下来，游戏开始，用户不断向系统输入操作命令，进行多轮互动。
% 例如，在第一轮，用户对机器人下达指令：士兵从d2到d4。
% Coordinator根据用户的命令和游戏情景信息进行反应，将针对棋子（gadget）的操作告知Controller：
Subsequently, as the game started, the user continuously gave commands to the system. 
For instance, in the first round, the user inputted a command: "Move the pawn from d2 to d4". 
Next, the Coordinator responded to the command and informed the Controller of the gadget behavior of the pawn actuated by the user, as shown in the middle block in \ref{fig:design space example}b.
% Controller agent接收到coordinator生成的关于棋子的操作后，进一步生成扮演士兵的robotic gadget的行动序列，用于控制该机器人的移动：
The Controlled then received the gadget behavior of the pawn, along with the actual location data (shown in \ref{fig:design space example}c) of the robot that represented the pawn gadget, proceeding to generate the corresponding action sequence for the pawn robot through the CoT process.
The output action sequence of the pawn robot is shown in the right block in \ref{fig:design space example}b.
% 此后，服务器通过python程序，解码Controller Agent生成的行动序列，向对应机器人发送控制指令，最终实现robotic gadget的行动行为。
The server in the system then decoded the action sequence from the Controller, sent the motion commands to the pawn gadget, and then actuated the movement of the pawn gadget, shown in \ref{fig:design space example}d.

\subsection{Interactive Behavior Planning}
% 在使用SUI进行桌面游戏时，取决于不同的游戏场景，SUI通常需要控制机器人的运动、将游戏中的各种操作转化为有型交互行为[ 46 ]。
% 然而，由于这些交互行为的复杂性和特殊性，系统中的agent很可能无法单纯依赖3.3中的通用性的提示词实现这些有形交互行为。
% 为了应对这一挑战，我们通过为Controller Agent添加若干组附加提示词（add-on prompt）的方法，辅助其理解一些特殊游戏操作的能力，并为集群机器人生成对应的行动序列。
% 我们也follow 3.3中设计基础提示词所采用的few-shot learning的方法，为每一组附加提示词提供了可供agent学习的案例。
Generally, in the context of utilizing SUI for tabletop games, depending on the specific game scenario, SUI often needs to control the movements of robots and translate various game operations into tangible interactive behaviors\cite{HERMITS}. 
However, in practice, we have found that due to the complexity and specificity of various interactive behaviors, the agents in our system, particularly the Controller, may not be able to rely solely on the generic prompts to realize all of these tangible interactive behaviors.
To address this challenge, we have enhanced the Controller's capability by incorporating several sets of additional prompts (add-on prompts) following a one/few-shot learning scheme. This approach aids the Controller in comprehending certain specific robotic operations (abilities) applied to different contexts. We anticipate that these add-on prompts will optimize the system's ability to generate action sequences for Gadgets across various game scenarios.

To determine which interactive behaviors required the design of additional prompts, we referenced past work in SUI \cite{HERMITS,shapebots,holobots,sparkybot,pengyu,AeroRigUI,ASTEROIDS} and recruited five designers with a background in Human-Computer Interaction (HCI) and experience in tabletop games for an informal interview and brainstorming session. During this process, we identified four common types of interaction behavior related to Gadget operations in tabletop games, including Objective Actuation, Symbol Visualization, Non-verbal Expression, and Scene Interaction. After that, we designed additional prompts for the Controller based on each type of interaction behavior, as detailed below.

\subsubsection{Object Actuation}
% 在基于SUI的桌面游戏中，如果系统能够使用robotic gadget根据游戏操作自动地驱动桌面上的物体[holo]，则可以实现对游戏中道具的自动操作、或模拟玩家与AI对手或远程玩家的互动。
% 为此，我们为controller添加一组附加提示词，辅助其为集群机器人生成用于“驱动物体”的行动序列。
% 在提示词中，我们规定conroller关注机器人移动的速度和轨迹，以实现物体按规定轨迹的移动、并仿真物体的属性，例如重量。
% 我们也在提示词中给出了一个具体的示例，包括推动找到一个重箱子、并推动到指定位置的任务，以及controller预期生成的行动序列。
% 具体提示词见附录。
In tabletop games based on SUI, the ability of the system to automatically manipulate objects on the table using robotic gadgets \cite{holobots} enables the automation of game prop operations or simulates interactions between players and AI opponents or remote players. To this end, we added a set of additional prompts for the controller to assist in generating action sequences for our Gadgets to "actuate objects". In these prompts, we specify that the Controller should focus on the speed and trajectory of robot movement to achieve object movement along a prescribed path and simulate the properties of objects, such as weight. We also included a specific example in the prompts, which involves the task of pushing a heavy box to a designated location and the expected action sequence generated by the Controller.

\begin{figure}[!htbp]
    \centering
    \includegraphics[width=\linewidth]{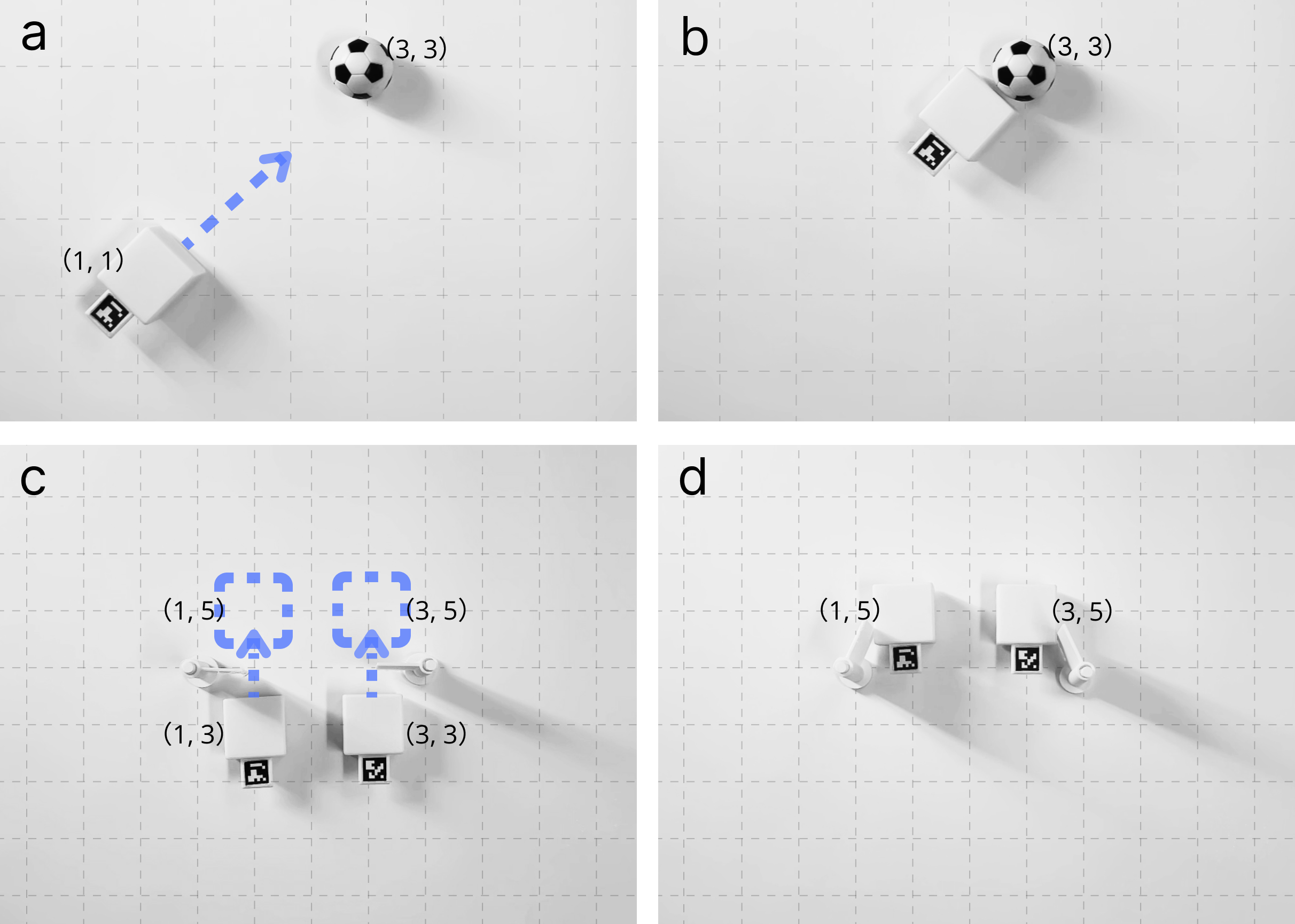}
    \caption{The gadget starts at position (1,1) and needs to kick a light plastic soccer ball located at (3,3). a) Gadget: Moves northeast at speed 3 to approach the soccer ball's location b)Kicks the soccer ball, simulating the action by moving east to (3,3). c-d) Two gadgets open a door together by moving towards it.}
    \label{fig:actuation}
\end{figure}

% 在向controller录入了驱动物体的提示词后，我们首先测试了使用controller驱动机器人推动重物场景的效果。
% 例如，规定两台机器人初始分别位于(1，1)和（1，7），gadget的行为需要他们一起把位于（7，7）（7，8）上的两扇很重的门推开。
% 机器人运动的效果以及controller生成的行动序列如图？。
After incorporating the add-on prompt for object actuation into the Controller's original prompt, we test the after-modified effectiveness by using the Controller to actuate the Gadgets in scenarios involving pushing heavy objects. 
We tested the effectiveness of using the controller to actuate the Gadgets in pushing light objects. For instance, it was stipulated that the Gadgets start at a position (1, 1) and need to kick a light plastic soccer ball located at (3, 3). The outcome of the Gadget's movement and the action sequences generated by the controller are depicted in Figure \ref{fig:actuation}a-b.

Also, in another example, two gadgets are initially located at (1, 1) and (3, 1), and we require the gadgets to push the two very heavy doors to (1, 4) (3, 4) from (1, 3) and (3, 3).
The demonstration of the movements of the Gadgets and the action sequence generated by the Controller is shown in Figure \ref{fig:actuation}c-d.

% 我们也测试了使用controller驱动机器人推动轻物体的效果。
% 例如，规定机器人初始位于(1，1)，并需要把停在（7，7）的足球踢开。
% 机器人运动的效果以及controller生成的行动序列如图？。

\subsubsection{Symbol Visualization}
% 在基于SUI的桌面游戏中，如果系统能够使用robotic gadget将某些图形符号进行可视化，则可以为游戏增加文字或图形信息的呈现能力 [ 74 ]。
% 为此，我们为controller添加一组附加提示词，辅助其为集群机器人生成用于“信息呈现”的行动序列。
% 具体地，我们设计了“描绘”和“组合”两种图形符号可视化的方法，以及每种方法对应的提示词。
% 为了使可视化功能的表现更加一致(consistent)，我们添加了一些规则，例如不能上下颠倒、英文字母为大写字母等。
% 我们也在提示词中分别为两种可视化方法提供了一个示例，包括可视化“HCI”三个字母的自然语言任务描述，以及controller预期生成的行动序列。
% 具体提示词见附录。
In SUI-based tabletop games, the system's ability to use robotic gadgets to visualize certain graphic symbols can enhance the presentation of textual or graphical information in the game [74]. For this purpose, we added a set of add-on prompts for the Controller to assist in generating action sequences for swarm robots (the Gadgets) for Symbol Visualization.
Specifically, we designed two methods for visualizing graphic symbols: "trajectory tracing" and "robotic formation," along with corresponding prompts for each method. To make this visualization function more consistent, we also added certain rules, such as "not inverting the symbols vertically" and "using uppercase letters for English alphabets". We also provided an example for each visualization method in the add-on prompt, including a description of a natural language task for visualizing the letters "HCI," along with the expected action sequence the Controller should generate. Detailed prompts are provided in the Supplementary Material.

\begin{figure}[!htbp]
    \centering
    \includegraphics[width=\linewidth]{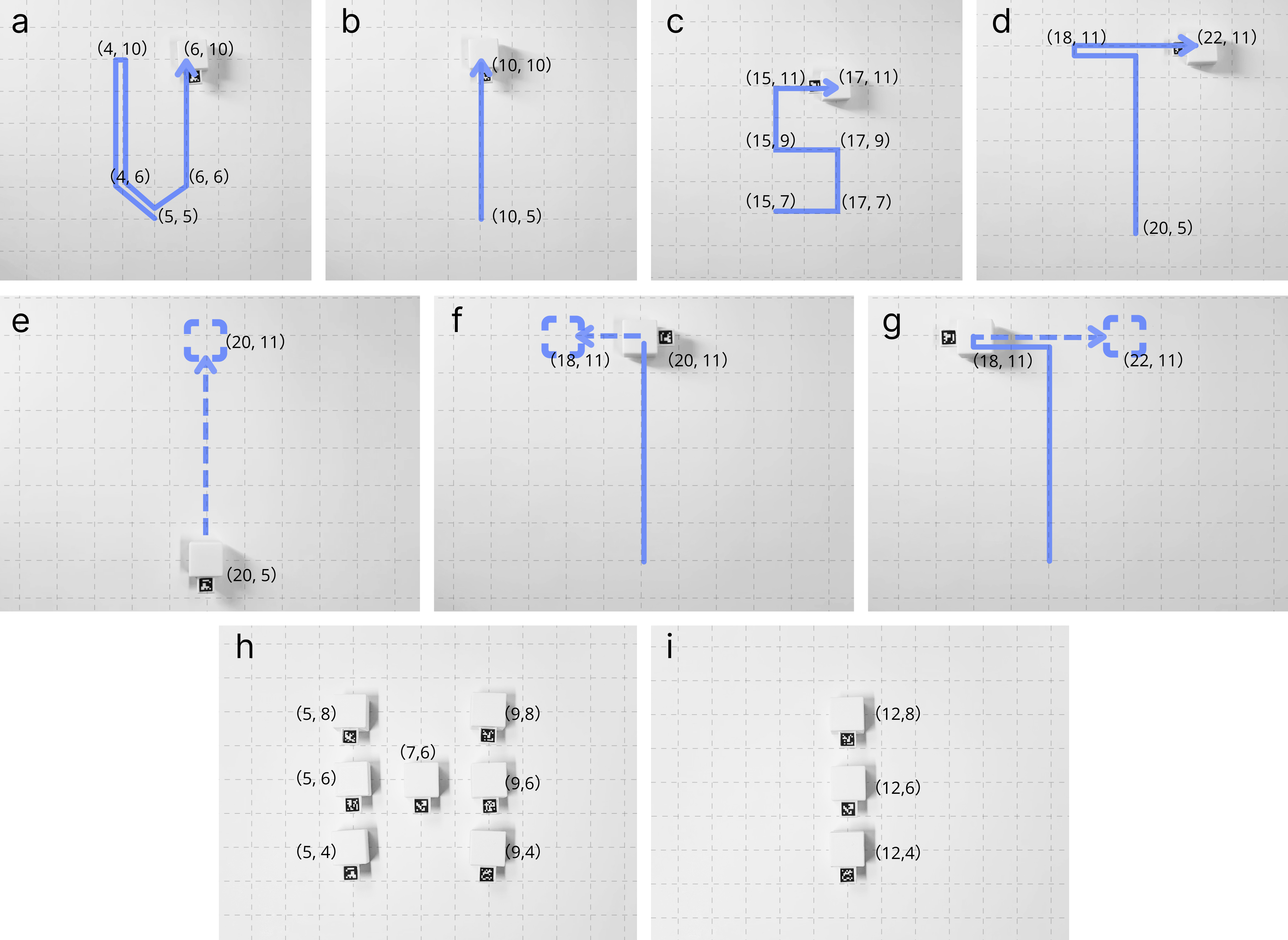}
    \caption{The gadgets were asked to present the word 'UIST' or 'HI'. a-d) Agents generated trajectories of four gadgets, which were arranged to form 'U', 'I', 'S'. e-f) A detailed demonstration of how one of the gadgets presents the letter "T" through its movement trajectory. h-i) Agents make each letter composed of multiple gadgets, using 7 and 3 gadgets to form the letters 'H' and 'I' respectively.}
    \label{fig:xinxichengxian}
\end{figure}

% 在向controller录入了信息呈现的提示词后，我们测试了使用controller驱动若干机器人可视化“UIST”的效果。
% 对于“描绘”方法，controller成功判定使用四台gadget，并为每台gadget生成对应的移动轨迹，轨迹图案以及形成轨迹的行动序列如图？。
After incorporating the add-on prompt for symbol visualization into the Controller, we tested its effectiveness by using the Controller to drive certain Gadgets in visualizing "UIST." For the "tracing" method, the Controller successfully determined to utilize four Gadgets and generated appropriate trajectories of movement for each Gadget. The trajectory patterns and the action sequences that formed these trajectories are illustrated in Figure \ref{fig:xinxichengxian}a-g.

% 对于“组合方法，controller则判定使用多台机器人分别组成"U"、"I"、"S"、"T"四个字母的形状，机器人组合字母的效果以及使用的行动序列如图？
For the "formation" method, the Controller determined to utilize multiple robots to separately form the shapes of the letters "H" and "I." The effectiveness of the robotic forming these letters and the action sequences used are illustrated in Figure \ref{fig:xinxichengxian}h-i.

% gadgets被要求呈现出"UIST"或”HI“的字样。a-d）Agents生成四个gadgets的行动轨迹，它们的轨迹分别组成”U“”I“”S“”T“。e-f）详细展示了其中一个gadget如何通过行动轨迹呈现字母”T“。h-i）Agents使每个字母由多个gadgets组成，分别使用7个和3个gadgets组成字母”H“和”I“

\subsubsection{Non-Verbal Expression}
% 在基于SUI的桌面游戏中，如果系统能够使用robotic gadget进行非言语表达，例如表现角色的情绪等，则能极大增强桌面游戏的叙事表现力[PENGYU]。
% 为此，我们为controller添加一组附加提示词，辅助其为集群机器人生成用于“非言语表达”的行动序列。
% 具体地，我们设计了“情绪表达”和“社交行为”两种非言语表达的方法，以及每种方法对应的提示词。我们也在提示词中分别为两种可视化方法提供了一个示例，以及controller预期生成的行动序列。两个示例包括伤心情绪、和两辆车之间的问候。
% 具体提示词见附录。
In SUI-based tabletop games, the ability of the system to use robotic gadgets for non-verbal expressions, such as displaying characters' emotions, can significantly enhance the narrative expressiveness of tabletop games \cite{pengyu}. To this end, we incorporate a set of add-on prompts for the Controller to assist in generating action sequences for swarm robots for "non-verbal expression." Specifically, we designed two methods of non-verbal expression: "mood expression" and "social expression," along with corresponding prompts for each method. We also provided an example for each method in the prompts, along with the expected action sequences the Controller should generate. The examples include asking the Gadget to express sadness and a greeting between two Gadgets. Related prompts are detailed in the Supplementary Material.

\begin{figure}[!htbp]
    \centering
    \includegraphics[width=\linewidth]{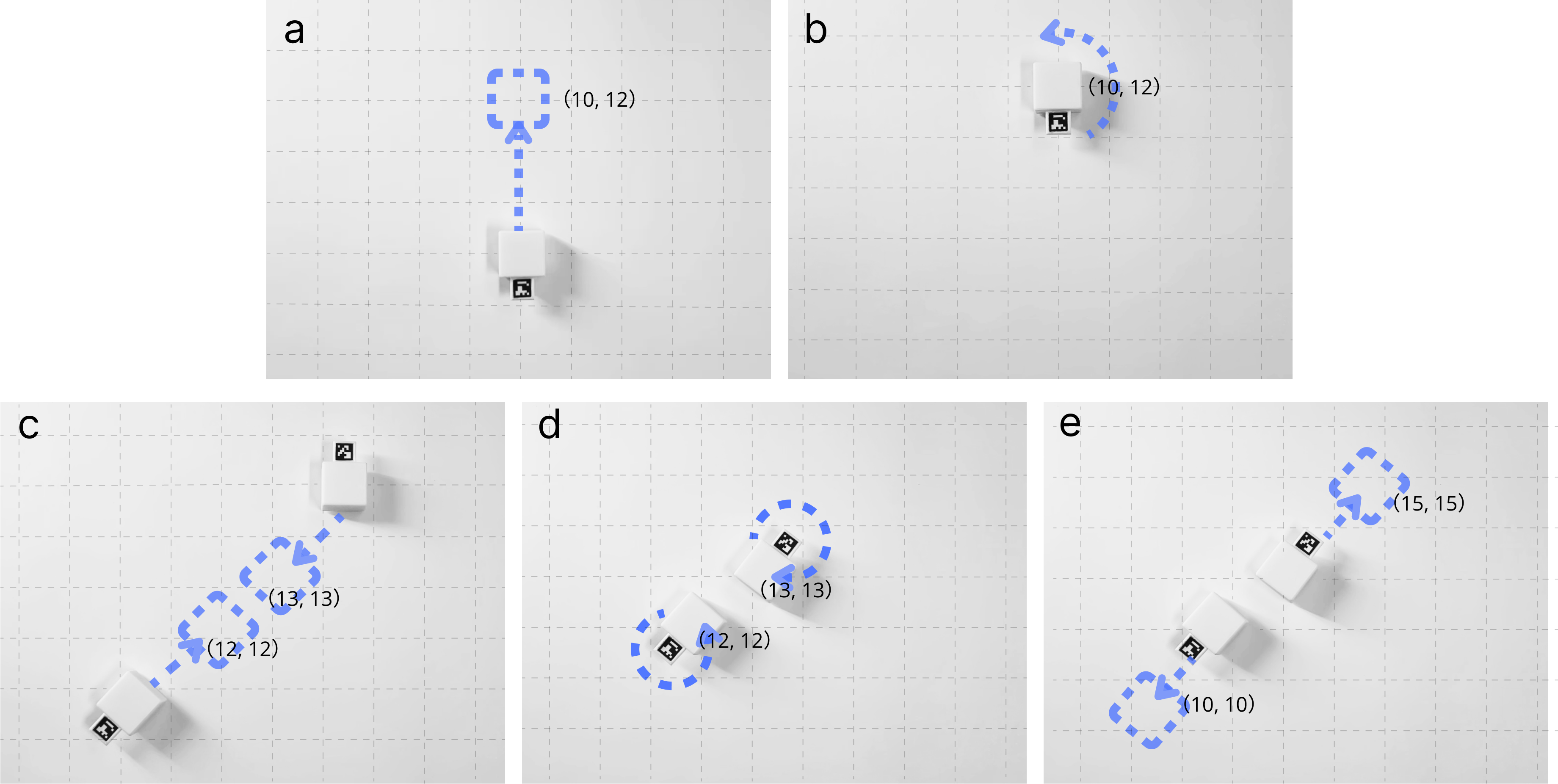}
    \caption{a-b) A single gadget expresses excitement, the gadgets move upward a little and rotate for a circle. c-e) Two gadgets express an argument, they move toward each other, each rotates for a circle during the argument, and each retreats when the argument is over.}
    \label{fig:non-verbal}
\end{figure}

% 在向controller录入了信息呈现的提示词后，我们首先测试了使用controller驱动单个机器人表达兴奋程序的效果，其表现以及对应的行动序列如图？。
After incorporating the add-on prompt for non-verbal expression into the Controller, we first tested the effectiveness of using the controller to drive a single Gadget to express excitement. The demonstration of this expression and the corresponding action sequences are illustrated in the Figure \ref{fig:non-verbal}a-b.

% 我们其次测试了使用controller驱动两个机器人表达争执社交行为的效果，其表现以及对应的行动序列如图？。
Next, we tested the effectiveness of using the controller to drive two Gadgets to express a disputing social behavior. The demonstration of this expression and the corresponding action sequences are illustrated in the Figure \ref{fig:non-verbal}c-e.

% a-b) controller驱动单个gadget表达兴奋,gadget向上小幅度移动并旋转一周 c-e)controller驱动两个gadgets表达争执，它们分别走向对方，争吵时各自旋转一周，结束争吵后各自撤退

\subsubsection{Scene Interaction}
% 在基于SUI的桌面游戏中，当希望robotic gadget在场景中进行storytelling等行为时，gadget的行为可能需要与地图环境进行互动[sparkybot]。
% 为此，我们为controller添加一组附加提示词，辅助其为集群机器人生成用于“场景交互”的行动序列。
% 具体地，我们在提示词中要求robotic gadget在行动规划时具备**全局视野**，考虑地图附加信息、场景道具、甚至是其它机器人位置或状态的因素等。
% 我们也在提示词中添加了一个示例，包括一个机器人绕过一条具有特定长度的障碍物，到达障碍物另一侧的例子。
% 具体提示词见附录。
% 在向controller录入了信息呈现的提示词后，我们测试了使用controller驱动一个机器人绕过由另外三个机器人组成的障碍物的效果，其表现以及对应的行动序列如图？。
In SUI-based tabletop games, as it is expected that a robotic gadget should be able to perform actions such as storytelling within a scene, the gadget's behavior may need to interact with the settings of map environment \cite{sparkybot}. To this end, we incorporate a set of add-on prompts for the Controller to assist in generating action sequences for swarm robots for "scene interaction." Specifically, we proposed that the Controller possesses a "global perspective" in the Gadget's action planning, taking into account additional map information, scene props, and even the positions or statuses of other Gadgets among other factors. We also provided an example for each method in the prompts, along with the expected action sequences the Controller should generate. The add-on prompt includes an example demonstrating a Gadget navigating around an obstacle of a certain length to reach the other side of the obstacle. Related prompts are specified in the Supplementary Material.

\begin{figure}[!htbp]
    \centering
    \includegraphics[width=\linewidth]{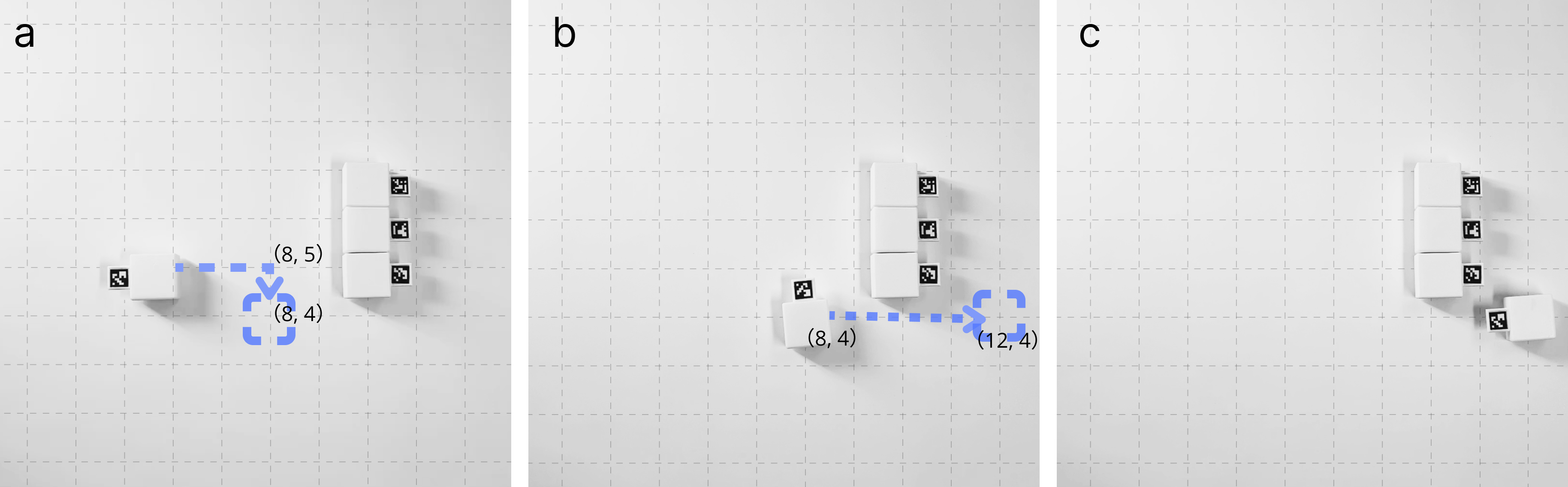}
    \caption{The gadget goes around a wall consisting of three other gadgets, approaching the obstacle before moving to the south side of the obstacle and moving east to get over the obstacle.}
    \label{fig:scene}
\end{figure}

After incorporating the add-on prompt for scene interaction into the Controller, we tested it with an example of navigating a Gadget around an obstacle formed by three other Gadgets. The demonstration of this expression and the corresponding action sequences are illustrated in Figure \ref{fig:scene}.

% gadget绕过由另外三个gadget组成的一堵墙，先接近障碍物，再绕到障碍的南边，向东移动以越过障碍物

\subsection{Interaction Relationship Planning}
% 在一项涉及多轮交互的游戏过程中，Agent通常需要考虑玩家与其交互的关系结构，通常包括：Apprentice，Competitor，Teammate或Designer[relationship]。
% 为了让Agent能够理解游戏中的人机交互关系与交互行为生成之间的关系，我们为controller agent 添加附加提示词（add-on prompt），以理解这些不同的交互关系。
Within the multiple interaction rounds in a game, it is crucial for agents to reflect upon the relational structure underpinning their engagements with players. This encompasses scenarios where agents may either confront or collaborate with players, or independently modulate their responses in alignment with the unfolding narrative, thereby contributing to the thematic ambiance. Drawing from this premise, researchers' investigations have distilled the potential stances an agent might assume into four distinct categories: Apprentice, Competitor, Teammate, or Designer \cite{relationship}.
To enable agents to grasp the relationship between human-computer interaction relationships and the generation of interactive behaviors within the game, we have augmented the Controller with the additional prompt (add-on prompt) to understand the varying interaction relationships.

\subsubsection{Apprentice}
% 在大多数桌游场景下，gadget通常会扮演玩家的化身或受玩家控制的道具[DND]。
% 如果系统能够作为“学徒”，根据用户提出的任意意见，自动、准确地调整robotic gadget的行为，那么就能更好地实现个性化的交互行为。
% 为此，我们为controller建立了一组附加提示词，让其 refer to and adjust action planning as much as possible according to the user's guidance ，具体提示词见附录。
Within the context of most tabletop gaming environments, Gadgets in our system typically embody the player's avatar or operate as player-controlled entities \footnote{http://scruffygrognard.com/}.
We believe that the ability of the system to act as an "apprentice," autonomously and precisely modifying the actions of robotic gadgets in response to users' suggestions, would significantly enhance personalized interactive behaviors.
To achieve this, we established a set of additional prompts for the Controller, enabling it to refer to and adjust its action planning as much as possible according to the user's guidance. The specific prompts are detailed in the Supplementary Material.

% 在向controller录入了关于学徒的提示词后，我们以“提速”的意见为例进行测试。
% 例如，当gadget行为仅要求一台机器人从A(5,5)运动到B(15,15)，controller生成的行动序列如下。
% 注意到，gadget目前的移动速度为2。
After incorporating the add-on prompt for "apprentice" into the Controller, we tested it with an example of requesting to "speed up Gadgets". The following is an example of an action sequence generated by the Controller controlling a Gadget to move from (5,5) to (10,10) on the grid map. The translation (movement) speed is set to 2.

\begin{figure}[!htbp]
    \centering
    \includegraphics[width=\linewidth]{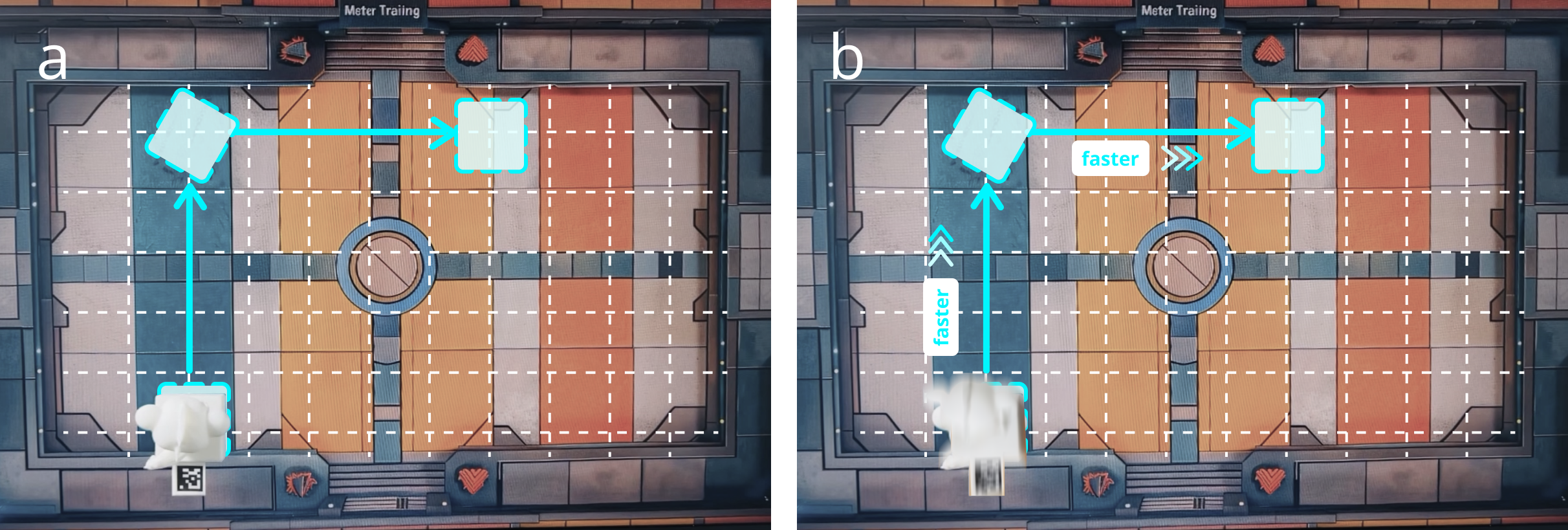}
    \caption{The gadget was asked to move from (5, 5) to (10, 10), and after being asked to speed up, the gadget moved faster.}
    \label{fig: Apprentice}
\end{figure}

% 接着，如果gadget的行为要求**更快些**地移动，controller则生成如下行动序列：
% 可以看到，controller随即调整了gadget的速度等级到最快，即速度等级3。
Following this, if we then request the Gadget to move faster, the Controller updates and generates the following sequence of actions as Figure \ref{fig: Apprentice}

% gadget被要求从（5，5）移动到（10，10），在被要求提速后，gadget移动速度变快

\subsubsection{Opponent}
% 许多桌面游戏存在非玩家控制的对手的角色，与玩家进行对抗活动。
% 如果系统能够理解“对手”关系，自动建立这些非玩家控制的对手角色，并使用gadget将互动实体化，那么就能够丰富桌面游戏的对抗性和游戏体验。
% 为此，我们为controller建立了一组附加提示词，帮助其生成并扮演单一或多个对手进行对抗或比赛的角色。
Many tabletop games feature the roles of opponents not controlled by players, engaging in competitive activities with the players. We believe it can enrich the gaming experience of tabletop games if our system comprehends the concept of "opponent", establishes non-player-controlled opponent roles, and uses the Gadgets to materialize their interactions. To facilitate this, we have established a set of additional prompts for the Controller, helping it generate and enact the roles of single or multiple opponents for competition or matches in games.

% 在提示词中，我们made planning[https://openreview.net/pdf?id=cMDMRBe1TKs] about controller在生成Opponent角色时所应遵循的原则和duties，例如……。并规定了controller如何分析玩家行为并动态制定富有挑战性的交互策略的机制，例如……
% 具体提示词见附录。
We specify in the prompt the planning about the principles and duties that the Controller should adhere to when generating "Opponent" roles \cite{raman2022planning}. For example, we set the goal of the opponent "... to challenge the opponent characters through strategy and decision-making while keeping the game fair and enjoyable." Furthermore, we have defined a mechanism for the Controller how to analyze player behavior and dynamically formulate challenging interaction strategies. For example, we specify in the prompt that the Controller "... formulate challenging strategies based on the current state of the game and the behaviors of opponent characters." The prompts are specified in the Supplementary Material.

\begin{figure}[!htbp]
    \centering
    \includegraphics[width=\linewidth]{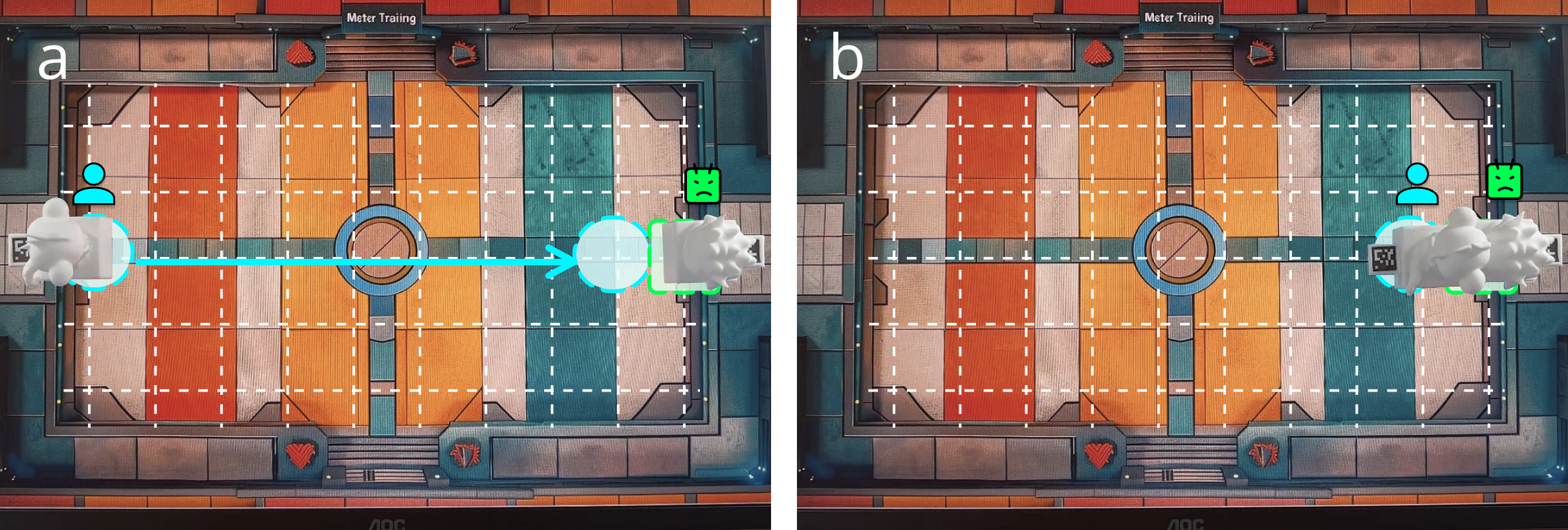}
    \caption{Monster1 is asked to attack Monster2, and Monster2 responds being attacked.}
    \label{fig:opponent}
\end{figure}

% 在向controller录入了关于“对手”的提示词后，我们测试了使用gadget进行对战行为的效果。
% 其中，一台gadget扮演用户控制的皮卡丘，并规定另一台gadget由系统控制、扮演杰尼龟作为对手。
% 当触发皮卡丘的电击攻击行为时，controller输出二者对战的行动序列如下，注意到……：
After incorporating the add-on prompt for "opponent" into the Controller, we tested it with an example of using Gadgets for combat behavior. In this example, one gadget embodies the role of a Monster1 commanded by the user (with action sequences generated by the Controller), while another gadget was controlled by the system, acting as an opponent Monster2.
After the Monster1 is commanded to use Thunderbolt to attack, the Controller generates the action sequence of their battle as Figure \ref{fig:opponent}.

% Monster1被要求攻击Monster2，而Monster2作出被攻击的反应

\subsubsection{Teammate}
% 与对手相似，在桌面游戏中，也存在许多非玩家控制的对手的角色，在游戏中支援玩家，共同完成某件任务或对抗对手。
% 如果系统能够理解“队友”关系，自动建立这些非玩家控制的队友角色，并使用gadget将互动实体化，那么就能够丰富游戏体验。
% 为此，我们为controller建立了一组附加提示词，帮助其生成并扮演单一或多个队友角色。
% 在提示词中，我们 also made planning[https://openreview.net/pdf?id=cMDMRBe1TKs] about controller 生成Teammate角色时所应遵循的原则和duties，例如……。并且，我们规定了controller如何通过有效的协作和策略支持队友角色，例如……，以确保整个团队的成功和游戏的乐趣。
% 具体提示词见附录。
Similar to opponents, in tabletop gaming, there exist numerous non-player controlled characters that support the player by collaborating to accomplish tasks or combat adversaries. Enriching the gaming experience becomes more feasible if our system can comprehend the "teammate (ally)" relationship, autonomously establish these non-player controlled teammate roles, and materialize interactions using the Gadgets. To achieve this, we have also developed a set of add-on prompts for the Controller, fostering it in generating and embodying one or multiple teammate roles. We specify in the prompt the planning about the principles and duties that the Controller should adhere to when generating "Teammate" roles \cite{raman2022planning}. For example, we set a goal for the teammate "... is to support teammate characters through effective collaboration and strategy, ensuring the success of the entire team and the enjoyment of the game..." Furthermore, we have established guidelines in the prompts for how Controllers can support teammate roles through effective collaboration and strategic planning, such as "providing necessary support to help teammate characters overcome challenges while ensuring not to overshadow them, maintaining the game's balance and interest.", to ensure the success of the entire team and enhance the enjoyment of the game. The prompts are specified in the Supplementary Material.

\begin{figure}[!htbp]
    \centering
    \includegraphics[width=\linewidth]{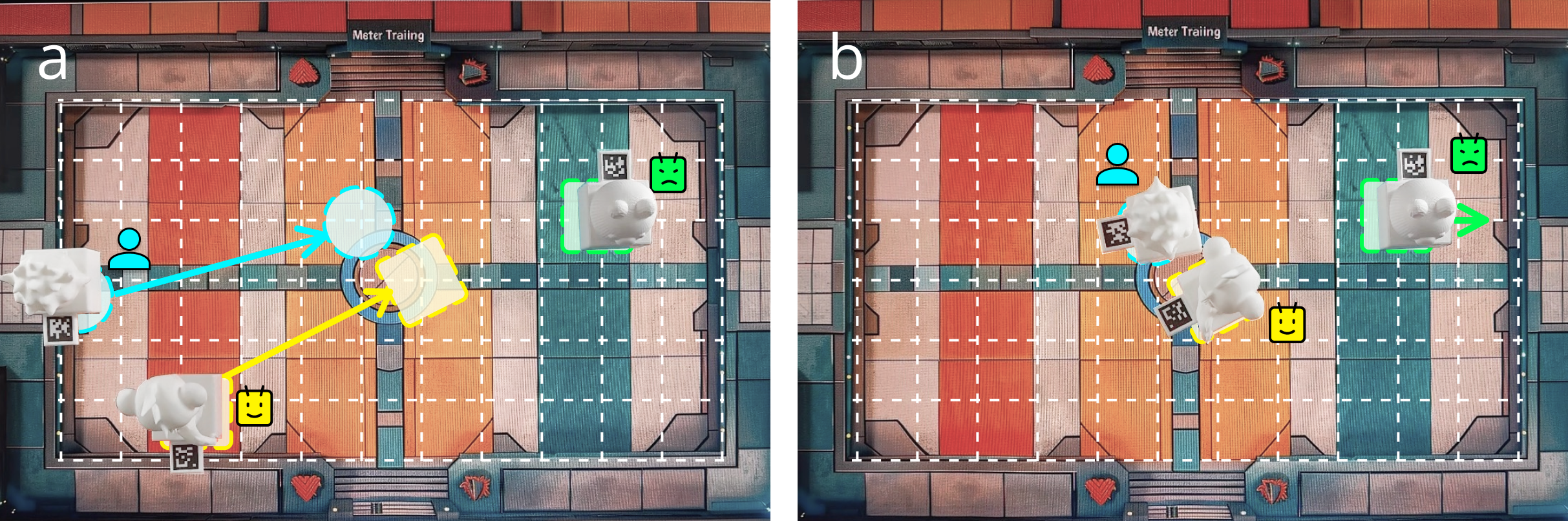}
    \caption{Gadgets that can generate collaborative behavior act as a teammate of the user and attack the enemy together.}
    \label{fig:teammates}
\end{figure}

% 在向controller录入了关于“队友”的提示词后，我们测试了使用gadget进行协同作战行为的效果。
% 其中，一台gadget扮演用户控制的皮卡丘，并规定另两台gadget由系统控制、分别扮演杰尼龟队友、与超梦对手。
% 当触发皮卡丘的电击攻击行为时，controller输出皮卡丘与杰尼龟协同作战的行动序列如下，注意到……：
After incorporating the add-on prompt for "teammate" into the Controller, we tested it with an example of using Gadgets for collaborative combat behavior as Figure \ref{fig:teammates}. In this scenario, one Gadget embodies the role of a Monster1 commanded by the user (with action sequences generated by the Controller), while two other Gadgets, generated and governed by the Controller, take on the roles of a teammate Monster2 and an opponent Monster3, respectively.

% 能产生协同行为的gadget作为用户的队友，共同攻击敌人。

\subsubsection{Designer}
% 在桌面游戏中，还可能存在其他辅助剧情的角色，例如NPC或道具。
% 例如，当玩家和队友合伙打赢了对手时，可以设置一些村民NPC来庆祝胜利。
% 如果系统能够作为“设计师”，根据游戏剧情自动生成这些角色，并使用gadget将他们的行为实体化，那么则能提升游戏的戏剧化效果与体验。
% 为此，我们为controller建立了一组附加提示词，让其……，具体提示词见附录。
In tabletop games, more roles that serve to enhance the narrative, such as NPCs (Non-Player Characters) or props, may also be present. For instance, upon the players and their teammates defeating adversaries, villager NPCs can be configured to celebrate players' victory. If the Controller can act as a "designer," generating these roles based on the game's narrative and materializing their actions through gadgets, then it would elevate the dramatic effect and overall experience of the game. To facilitate this, we have developed a set of add-on prompts for the Controller, enabling it to "... spontaneously generate new characters, NPCs, items, or plots, as well as the corresponding robotic action sequences, to advance the game storyline." Related prompts are detailed in the Supplementary Material.

\begin{figure}[!htbp]
    \centering
    \includegraphics[width=\linewidth]{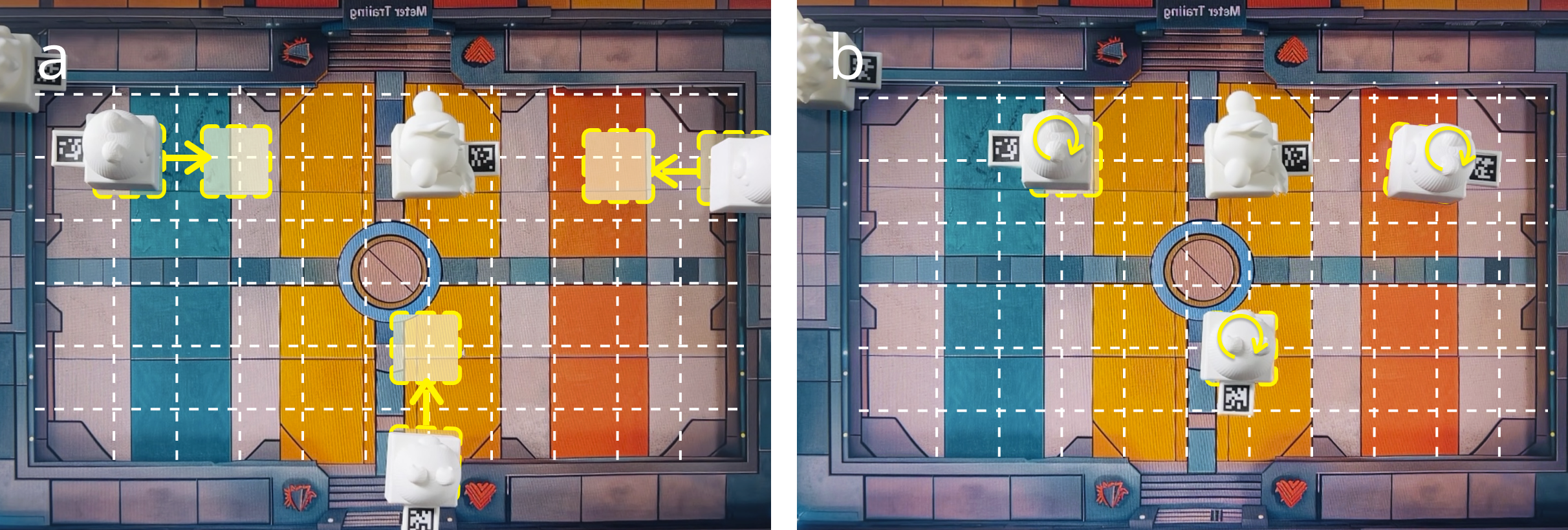}
    \caption{Three new NPCs generated by the agent celebrate the user's victory together, they surround the player, each taking a step forward and expressing excitement by spinning.}
    \label{fig:designer}
\end{figure}

% 在向controller录入了关于“设计师”的提示词后，我们测试了一次对战后胜利的关键剧情。
% 具体地，当用户控制的皮卡丘和杰尼龟战胜了超梦后，controller自发地生成了村民NPC庆祝的剧情，并调用3个新的robotic gadget进行庆贺行为：
After incorporating the add-on prompt for "designer" into the Controller, we tested it with an example focusing on a pivotal narrative moment following a battle victory. Specifically, when the user-controlled Monster1 and the teammate Monster2 defeated Monster3, the Controller generated a storyline in which villager NPCs appeared to celebrate the victory. To bring this celebratory behavior to life, the Controller then invoked three new Gadgets designed for this purpose as Figure \ref{fig:designer}.

% 由agent生成的三个新npc共同庆祝用户的胜利，它们围绕着玩家，各自向玩家方向前进一步，并通过旋转一周表达兴奋

\section{Example Application}
% 我们使用AI-Gadget Kit设计了几种不同的桌游，以展示其使用two-agent system以及以上八个add-on prompt（4个interaction Behavior和4个Interaction Relationship）的效果。
% 在玩每一种桌游时，我们均按照（）中的步骤，首先在AI-Gadget Kit输入游戏介绍和规则（见附录），声明所进行的游戏，其次根据游戏场景输入游戏命令，与gadget进行互动游戏。
We designed several different tabletop games using the AI-Gadget Kit to demonstrate its capabilities with a two-agent system and the effectiveness of the eight add-on prompts (comprising four Interaction Behaviors and four Interaction Relationships). For each game played, we followed the procedure outlined in Figure \ref{fig:design space example}, beginning with giving the game's introduction and rules to the AI-Gadget Kit (details provided in the Supplementary Material). This initial step involved declaring the game being played. Subsequently, based on the game scenario, we provide game commands to interact with the Gadgets, engaging in an interactive gameplay experience.

\subsection{Soccer-Ball-Shooting Game}
\begin{figure}[!htbp]
    \centering
    \includegraphics[width=\linewidth]{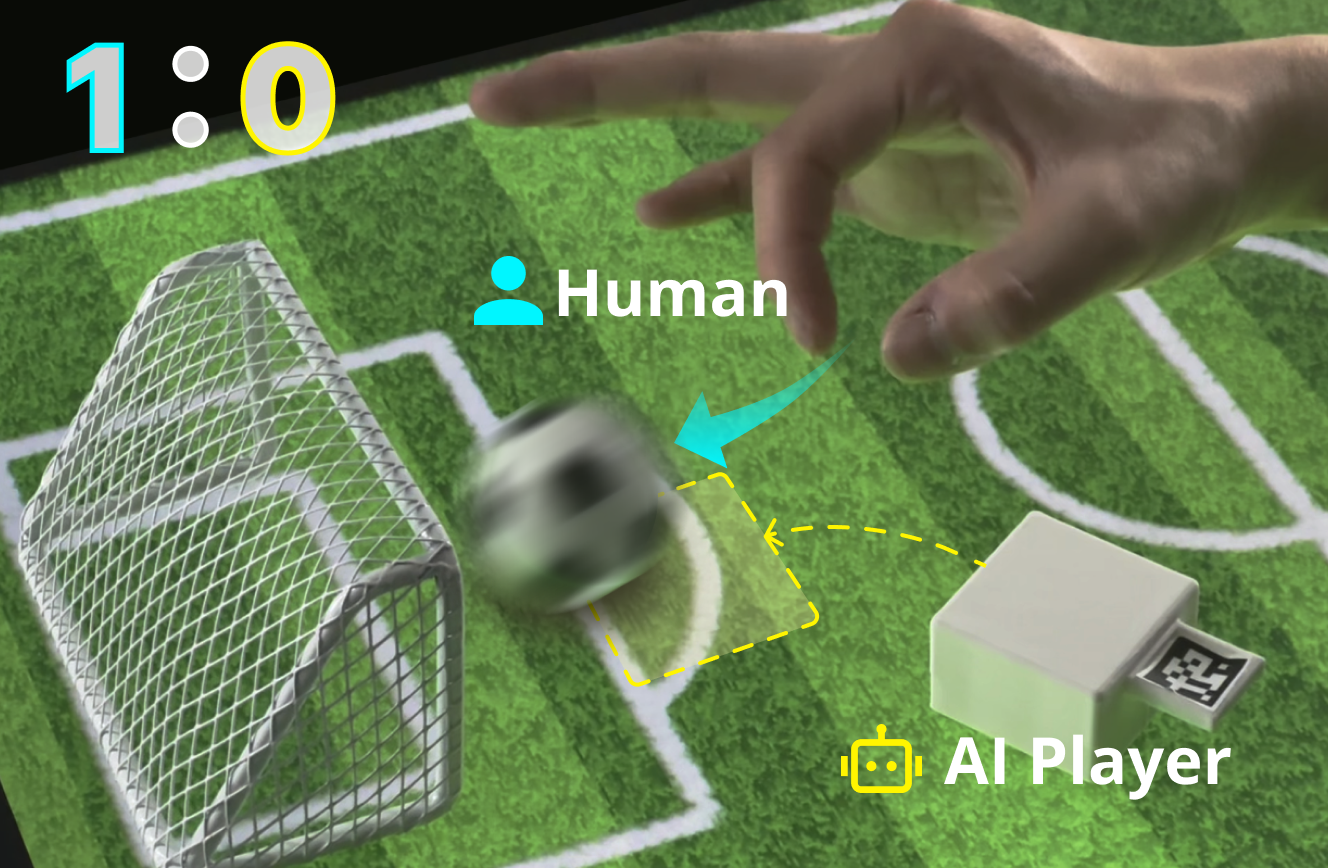}
    \caption{Human player and AI player play soccer games together.}
    \label{fig:soccer}
\end{figure}

% 我们用AI-Gadget kit实现了一个足球射门比赛，如图（）所示。
% 桌面上有一个位置固定的球门，玩家和robotic gadget在游戏中轮流进行足球射门，其中玩家用两根手指推动足球，机器人则通过撞击来完成踢足球的动作。
% 我们首先向系统输入射门比赛的游戏规则（见附录）。
% 在游戏中，Coordinator Agent理解游戏规则后，将在游戏中担任主持人和裁判身份，并让Controller调配一个对手 角色来控制一台机器人的行动。
% Kit将利用controller agent的“物体驱动”add-on prompt的能力，实现基于球门和足球位置的碰撞行为规划。
% 开始游戏后，当轮到机器人踢球的轮次时，our kit将根据足球距离球门的距离、球门相对足球的方位、推动足球所需的速度等因素生成机器人球员的行进方向和速度。
Using the AI-Gadget Kit, we implemented a soccer-ball-shooting game, as illustrated in the Figure \ref{fig:soccer}. The setup includes a stationary goal on the tabletop, with players and a Gadget taking turns to shoot the ball. Players propel the ball using two fingers, while the Gadget executes its shots by striking the ball.
We began by providing the game rules into the system (details in the Supplementary Material). Once the Coordinator comprehended the rules, it assumed the roles of both host and referee within the game. It then tasked the Controller with deploying an opponent role to control the action sequences of the Gadget.
The Kit shall leverage the capability through the "object actuation" add-on prompt of the Controller to simulate and execute collision planning based on the position of the goal and the ball. When it is the Gadget turn to take a shot, Controller generates the opponent's direction and speed by considering factors such as the distance of the ball from the goal, the ball's orientation relative to the goal, and the speed required to propel the ball.

% 图？展示了 该系统在该游戏情景下的表现效果。其中，robotic gadget作为对手，在玩家射门之后，自行瞄准发球点，向球移动，驱动足球进门。
Figure \ref{fig:soccer} demonstrates the performance of the system in this gaming scenario, highlighting the Gadget acting as the opponent. After the player takes a shot, the Gadget autonomously aims for the kickoff spot, moves towards the ball, and pushes it towards the goal.

% 人类玩家和AI玩家共同进行足球游戏

% 同时，agents还会会关注过去的比分，以增强游戏性和竞技性为目的调整策略。例如，在一局游戏中，当玩家连续多次没有得分，后续轮次中机器人的命中率也发生了下降，以增添玩家的游戏体验。
% （定性表达，体现关键meta-action, interaction behavior, relationship ）
Additionally, the Controller monitors past scores to adjust strategies to enhance gameplay and competitiveness. For instance, in a single game session, if the player fails to score consecutively over multiple attempts, the opponent's accuracy will be dynamically lowered in subsequent rounds. This adjustment is designed to enrich the users' gaming experience by maintaining a competitive balance and ensuring that the game remains engaging and challenging without becoming discouragingly difficult.

\subsection{Turn-Based Strategy Game}
\begin{figure}[!htbp]
    \centering
    \includegraphics[width=\linewidth]{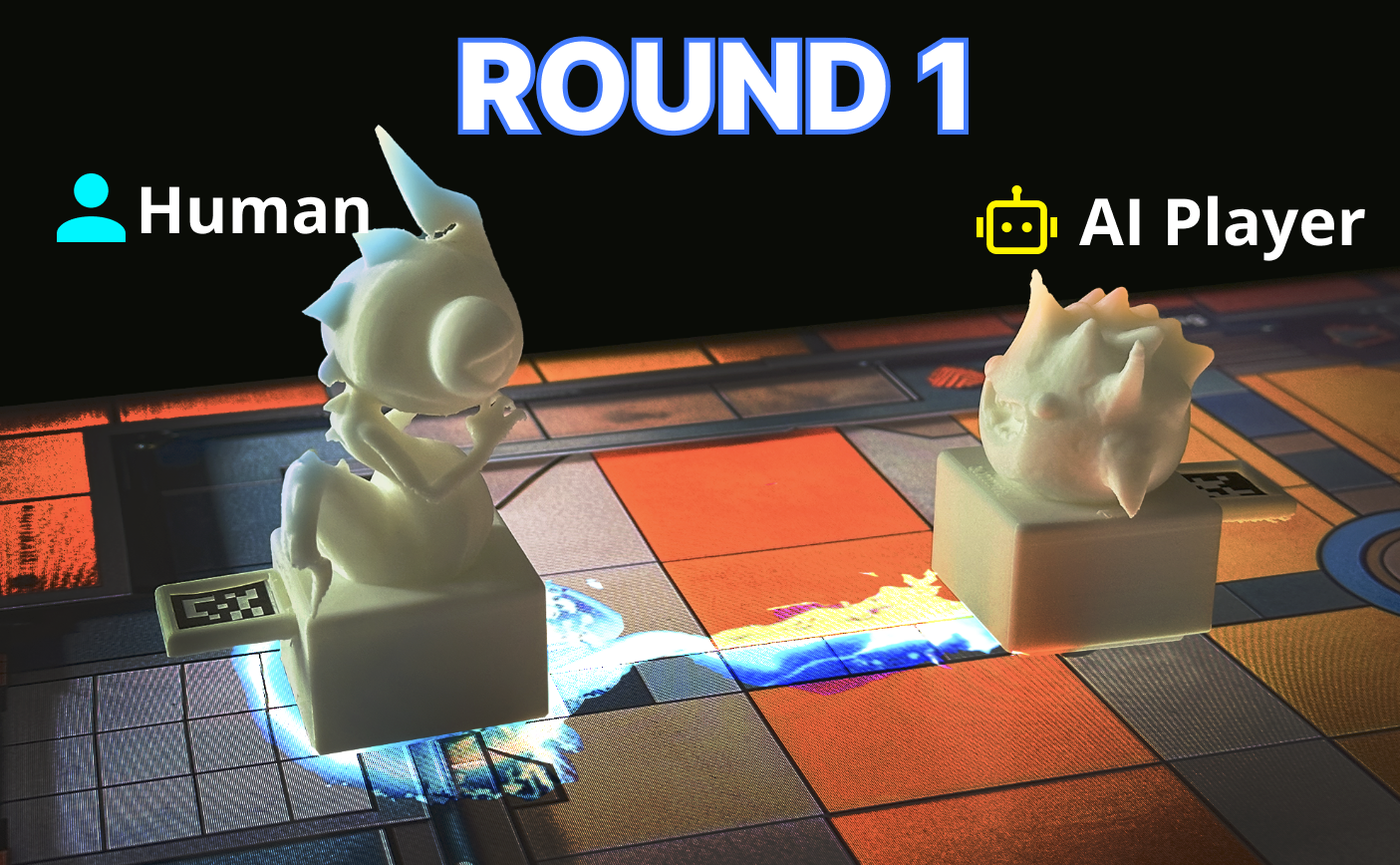}
    \caption{Robotic gadget plays turn-based strategy games with human players.}
    \label{fig:TBS}
\end{figure}

% 我们用AI-Gadget kit设计了一种回合制策略类游戏。
% 游戏中，玩家们都拥有自己的gadget角色，他们可以轮流释放指令，当且仅当轮到属于自己的回合才能够对角色进行操纵。
% 我们向系统输入the rule of Turn-Based Strategy game（具体见附录）。
% 在游戏中，Gadget能够支持玩家与玩家之间操控机器人的对战（PVP）、或与Gadget作为敌对方的对战（PVE）。
% 在游戏过程中，robotic gadget将扮演参与对战的角色，并根据各方玩家指定的攻击对象与释放的技能等内容生成机器人的行动指令。
% 依据上述游戏规则，Coordinator Agent会在交互中更新和传递游戏信息，如Gadget的状态、能力等信息。
% Controller agent会结合这些信息进行相应的行动序列生成。
% 在对战中，our kit主要基于学徒、对手以及非言语表达的add-on prompt，遵循玩家用户在对战过程中的交互指令（如攻击/防御、技能释放等）生成相应的交互行为。
% 如图？所示，当玩家要求自己的角色“用雷电的力量”攻击敌人时，kit将生成面向敌方机器人的攻击行动。
% 考虑到“雷电”所具备的特征，kit设计了攻击方的gadget模拟积蓄电量的倒退后猛冲的动作、以及作为对手的，受击gadget遭受电击后模拟触电的的原地摇摆动作。
% 此外，通过我们kit中的队友以及设计师的add-on prompt，TBS游戏还可以容纳更多的gadget进行对战或互动。这可以为现有DND类游戏中的战斗剧场进行补充[DND]。
Within the Turn-Based Strategy (TBS) game scenarios, the system is designed to support battles between players whose commands are embodied through Gadget controlling (Player vs. Player, or PVP) as well as battles where the Gadget itself acts as the opponent (Player vs. Environment, or PVE).
During gameplay, robotic gadgets assume roles as combatants, generating robotic action commands based on the attack targets and skills released as designated by the players. Following the game rules outlined, the Coordinator updates and transmits game information during interactions, such as the status and capabilities of the gadgets. The Controller then synthesizes this information to produce corresponding sequences of actions. In combat scenarios, our Kit primarily relies on an apprentice, opponent, and non-verbal expression add-on prompts. It adheres to the players' interaction commands (such as attack/defense, skill deployment, etc.) to generate corresponding interactive behaviors.
As illustrated in the figure\ref{fig:TBS}, when a player commands their character to "attack the enemy with the power of thunderbolt," the kit generates an attack action directed at the opposing Gadget.
Considering the characteristics of "thunderbolt," the kit is designed with actions for the attacking gadget that simulate the buildup of electrical charge followed by a swift charge forward, and for the opponent gadget, it simulates the motion of being electrocuted with on-the-spot swaying movements. Moreover, through the add-on prompts of teammates and designers within our kit, TBS games can accommodate a larger number of gadgets for combat or interaction. This can complement the combat theater in existing Dungeons and Dragons (D\&D) type games.

% 人类玩家与robotic gadget进行Turn-Based Strategy游戏

\subsection{Yes Or No?}
% 我们发现在很多游戏中存在由系统进行Yes or No回复的环节，如DND游戏中的能力判定、剧情中的问题回答等。
% 我们的kit可以帮助gadget在此类环节中对Yes和No进行更具趣味性的展示。
In many games, there are moments where a system provides Yes or No responses, such as ability checks and answering questions within the narrative in Dungeons and Dragons (D\&D) games. Our kit enhances the presentation of Yes and No answers in these instances, making them more engaging.

% 例如，我们设计了一个问答游戏（具体规则见附录）。游戏中，Coordinator Agent可以接受用户输入的问题并给出回答。
For instance, we have designed a quiz game\ref{fig:yesorno} (its rules are detailed in the Supplementary Material) where the Coordinator accepts questions raised by users and provides corresponding answers.

% 随后，具有设计师能力的Controller Agent会基于信息呈现的add-on prompt，通过控制Robotic Gadget在纸面上写出Y（Yes）或N（No）来对玩家的提问进行是或否的回答。
The Controller, equipped with designer capabilities, utilizes add-on prompts for symbol visualization to control a Gadget. This Gadget writes out Y (Yes) or N (No) on paper, responding affirmatively or negatively to the users' queries. 

% 为了消除笔尖与纸张间的摩擦力对robotic gadget运动的影响，实际场景中我们利用了两个机器人共同夹着一支笔进行绘制。
% 我们期望这样的设计能够给玩家带来更富有悬念和沉浸感的游戏体验。
To mitigate the impact of friction between the pen tip and paper on the robotic gadget's movement, we employed two gadgets to hold a pen jointly for drawing. This design is intended to offer users a more suspenseful and immersive gaming experience.

% Robotic gadgets通过驱动笔在纸上写字，对用户的提问做出yes或no的回答

\begin{figure}[!htbp]
    \centering
    \includegraphics[width=\linewidth]{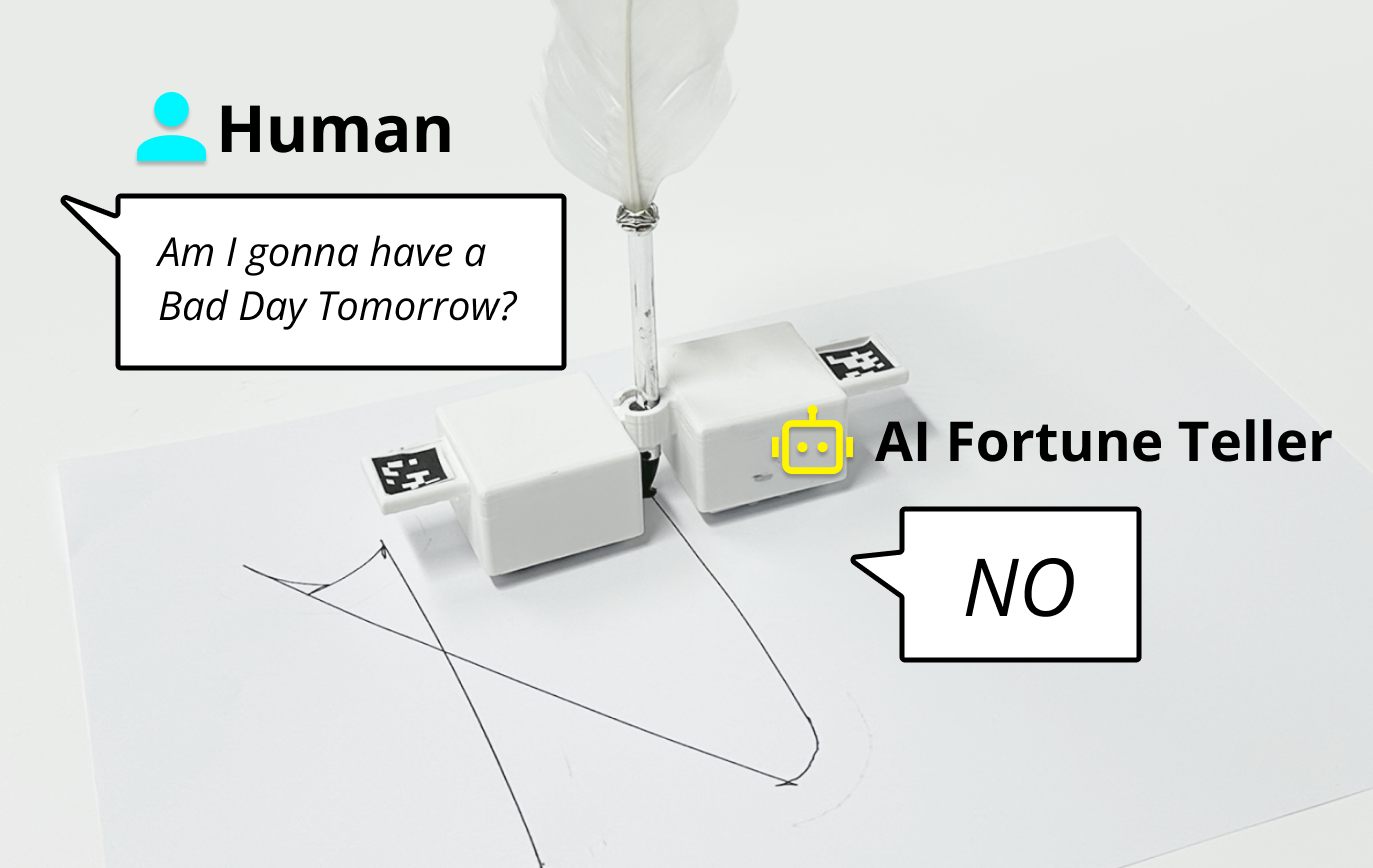}
    \caption{Robotic gadgets answer 'Yes' or 'No' to user questions by driving a pen to write on paper.}
    \label{fig:yesorno}
\end{figure}

\subsection{Improvisational Theater}
% 结合AI- Gadget，我们为用户设计了与机器人合作进行即兴戏剧表演的应用。即兴戏剧是一种没有固定剧本的团体戏剧形式，大部分表演内容都是演出过程中表演者的即兴创作结果。我们要求AI-Gadget遵循“Yes, and”的即兴戏剧表演核心原则，即在结合前一表演者表演内容的前提下，自由进行后续表演内容的生成。生成内容包括符合角色身份和剧情发展的角色语音、表达角色情绪状态的非言语情感表达行为、以及与当前表演场景信息产生的交互行为等。
Improvisational theater is a form of group drama without a fixed script, where most of the content is spontaneously created by the performers during the performance\cite{improvisational_theater}. Integrating AI-Gadget Kit, we have designed an application for users to perform improvisational theater in collaboration with the Gadgets. In this setup, Gadgets act as performers, with one of the Gadgets representing the user. We require the Gadgets to adhere to the core principle of improvisational theater, "Yes, and," which means freely generating subsequent performance content, including voice and action sequences, based on the previous performer's contribution and randomly selecting the next performer from among the users and other Gadgets. The Coordinator is responsible for assigning roles to all parties, recording and transmitting the content of the performance, and coordinating the turns of the performance.

\begin{figure}[!htbp]
    \centering
    \includegraphics[width=\linewidth]{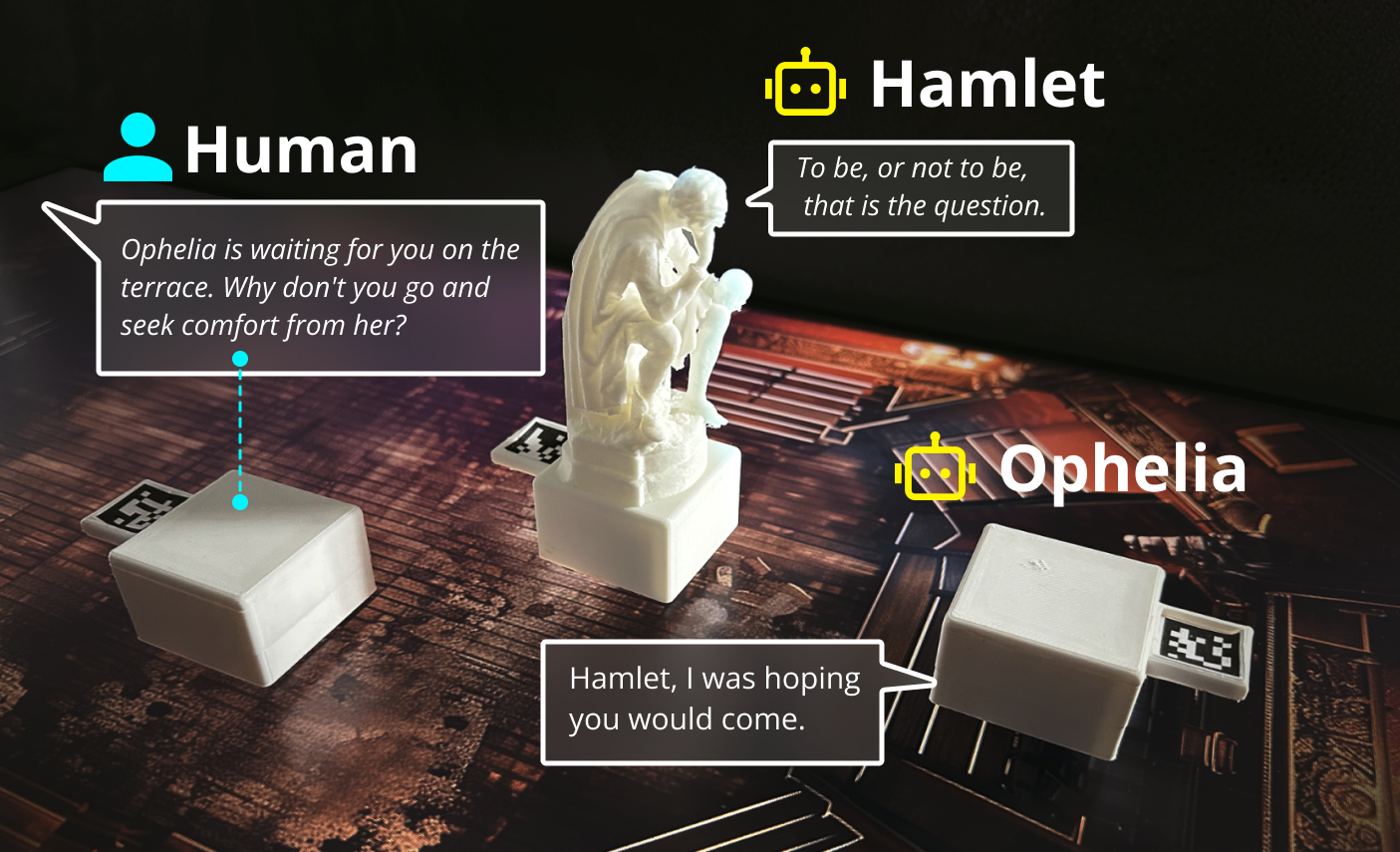}
    \caption{In Hamlet-themed improvisational theater, the gadget playing Hamlet flexibly reacts to the user's words outside of the script.}
    \label{fig:hamlet}
\end{figure}

% 为了在一定程度上控制戏剧的设定，我们测试了一个以戏剧《哈姆雷特》的世界观为背景的即兴戏剧。除了一个被预设为哈姆雷特扮演者的机器人，我们没有预设任何用户和机器人的角色身份。AI-Gadget首先为机器人集群生成了各自的身份，其中一部分遵循了戏剧原作中的人物，另外一些则是原创的角色。表演开始后，AI-Gadget将基于前一表演者的表演内容进行后续表演语音和动作的生成，并从用户和其他机器人中随机地选择下一段表演者。
After integrating non-verbal expression add-on prompts, the Controller is capable of generating dialogue that fits the characters' identities and the plot development, while also endowing robots with non-verbal "mood" expressions to convey the characters' emotional states. Moreover, by assigning specific scene information to different areas of the venue, the scene interaction capability add-on prompt enables the Controller to consider information such as the performance location when generating action sequences. In one of our improvisational theater tests set in the world of "Hamlet" (detailed game rules can be found in the Supplementary Material), our Kit initially assigns identities to a cluster of Gadgets(Figure \ref{fig:hamlet}), several following the original characters of the drama and others being original characters based on user commands.

% 表演开始后，Controller Agent会根据我们事先设置的表演场地信息，与用户进行配合，共同延续表演。例如，在饰演哈姆雷特的机器人说出“To be or not to be, that is a question.”时，与原著剧情完全不同地，用户会通过饰演Hamlet的朋友（原创角色）的机器人对他的低落情绪表达安慰，并告知他Ophelia正在露台上等候。根据用户输入的表演台词，Controller指挥代表用户的机器人通过非语言的行动序列表达情感。随后，Hamlet根据场地信息，移动到代表露台的区域，并与Ophelia进行了对话和表演。
Once the performance starts, the Controller coordinates with the user, taking into account the pre-set information of the performance site, to continue the act. For example, when the robot playing Hamlet utters "To be or not to be, that is the question," in complete contrast to the original story, the user, who is consoled by a robot as Hamlet's friend (an original character), expresses his comfort and informs Hamlet that Ophelia is waiting on the terrace. Based on the lines entered by the user for the performance, the Controller Agent directs the user's robot to express emotion through a non-verbal sequence of actions. Hamlet then moved to the area representing the terrace based on the venue information and engaged in a dialog and performance with Ophelia.

% 不难看出，AI-Gadget在表演场景中具有响应用户高自由度输入的能力。我们还测试了完全原创的故事背景，AI-Gadget在其中同样展示了对于即兴内容的接受和反馈能力。随着即兴戏剧这一艺术形式在想象力与创造力激发、心理健康和社会情感教育领域[Collaborative Emergence and Creativity][Creativity, Acceptance, and Psychological Well-Being][social anxiety]中的深入介入，结合更具专业性的即兴戏剧生成策略[Emergent Story Generation]，AI-Gadget有潜力为用户提供更加丰富的交互体验和情感体验。
It is not difficult to see that AI-Gadget has the ability to respond to high degree-of-freedom user input in performance scenarios. We also tested completely original story contexts in which AI-Gadget similarly demonstrated the ability to receive feedback on improvised content. As the art form of improvisational theater is gaining traction in the areas of imagination and creativity stimulation, mental health, and social-emotional education \cite{CollaborativeEmergenceandCreativity}\cite{CreativityAcceptanceandPsychologicalWellBeing}\cite{socialanxiety}, combined with a more specialized Emergent Story Generation strategy, AI-Gadget has the potential to provide users with a richer interactive and emotional experience.

% 在《哈姆雷特》主题的即兴戏剧中，扮演哈姆雷特的gadget灵活地对用户在剧本之外的语言作出反应。

\section{Discussion}

\subsection{Limitation and Future Works}
% 在本研究中，我们使用了Sony Toio系列机器人作为我们的Robotic Gadget。然而，由于Toio机器人的移动基于motor的差速控制，因此motor本身的性能将对机器人的行动表现产生影响，进而为AI- Gadget的表现带来的限制。例如，在实际测试过程中，Toio机器人在执行现有的移动指令时会产生随机偏差，这种情况会因为外壳增加附件时导致的重心偏移而更加明显。
In this study, we employed the Sony Toio robots as our Robotic Gadget. However, due to the Toio robots' movement being reliant on motor differential speed control, the performance of the motors themselves can impact the robots' action performance. This imposes limitations on the AI-Gadget Kit's functionality. For instance, in practice, the Toio robots exhibited random deviations when executing users' movement commands, a situation exacerbated by a shift in the center of gravity due to additional attachments to the casing.

% 为了应对上述因素给Gadget带来的挑战，我们正在考虑在系统架构上进行多个方面的优化。首先，更精确的机器人端的行进校正算法可以在机器人的移动表现上提供支持。对于Gadget预期的行动序列，结合摄像头中基于ArUCo的实时定位信息，机器人有能力在行动期间进行精确的闭环控制，实时对行动进行校正。此外，参考Swarm Haptics[]、Hermits[]等工作，对机器人进行机械结构的改造（如轮胎摩擦力的增加）可以被用来解决在驱动物体等使用场景下，Gadget的线性驱动能力和表现因为集群机器人的质量和速度而受限的问题。
To mitigate the challenges faced by the Gadget due to the aforementioned factors, we are considering several optimizations to the system architecture. Firstly, a more precise movement correction algorithm on the robot side could support the robots' movement performance. With the anticipated action sequences of the Gadget, coupled with real-time positioning information based on ArUCo from cameras, the robots are capable of precise closed-loop control during action, enabling real-time adjustments. Furthermore, inspired by works such as Swarm Haptics\cite{SwarmHaptics} and Hermits\cite{HERMITS}, mechanical modifications to the robots (like increasing tire friction) could address the limitations in the Gadget's linear driving capability and performance due to the mass and speed of swarm robots in scenarios such as object propulsion.

% 除去机器人行动能力方面的优化以外，我们同样意识到机器人端目前提供的meta-action不足以应对更广泛的复杂移动的需求，而对于Gadget而言，多元的移动方式（如曲线移动等）有望赋予机器人以更好的交互体验和表现力。因此，更多移动方式的封装和提供将协助Robotic Gadget在桌面游戏中进行具有复杂和丰富语义的交互表达。
Beyond optimizing the robots' movement capabilities, we also recognize that the current provision of meta-actions by the robot side is insufficient to meet the demand for a broader range of complex movements. For the Gadget, diverse movement modes (such as curved motion) are expected to improve the robots' interactive experience and expressiveness. Therefore, the encapsulation and provision of more movement modes will assist the Robotic Gadget in achieving interactions with complex and rich semantics in tabletop games.

% 在本研究中，在面对复杂的桌游场景时，例如你画我猜，agent所基于的LLM（即GPT-4）可能会因为添加的大量add-on prompt和过多的游戏回合，使其处理过长的上下文而遇到性能瓶颈。这会导致生成错误或者所谓的“幻觉”现象，即模型在理解或生成与游戏规则、情景描述相关的内容时产生误差。这种情况下，模型的输出可能需要额外的测试（e.g., The Needle In a Haystack Test）来验证其准确性和可靠性。
On the other hand, in this study, when confronted with complex board game scenarios, such as "Improvisational Theater," the LLM (i.e., GPT-4) which the agents rely on, may encounter performance bottlenecks due to the utilization of multiple add-ons prompts and numerous game rounds, leading to complications in processing lengthy contexts. This can result in inaccuracies and errors or the so-called "hallucination" phenomenon of LLM, where the model generates content with inaccuracies related to the game's rules and scenario descriptions. In such cases, the model's output may require additional testing (e.g., The Needle In a Haystack Test) to verify its accuracy and reliability.

% 为了应对这一挑战并提高模型在处理复杂场景下的表现，未来我们可以采取多种策略。首先，引入功能更强大的LLM是一个直接的解决方案。通过增强模型的理解能力和retrieval ability of long context，可以直接提高其在复杂交互场景下的表现。然而，此方案取决于外部技术的进步。
To address this challenge and enhance the agents' performance in complex scenarios, several strategies can be considered for the future. Firstly, introducing a more powerful LLM represents a direct solution. By enhancing the model's capability to understand and retrieve long contexts, we can directly improve agents' performance in complex interaction scenarios. However, this approach depends on external technological advancements.

% 此外，参考Autogen，我们可以引入更多细分的agent来处理特定的上下文问题。这些细分agent可以被设计为分别实现游戏规则生成、action sequence generation、事件后果分析等特定功能。细分的agent通过隔离特定的上下文，实现主要的agent间的交互。它们能够共同作用于提供更准确，有效和连贯的生成内容。例如，通过分发特定部分的上下文，某些agent可以专门负责追踪游戏状态和玩家动作的逻辑关系，而其他agent则可能负责生成与特定游戏环境或情景相符的描述和反应。
Furthermore, drawing inspiration from AutoGen\cite{autogen}, we can introduce multiple specialized agents to handle specific contextual challenges. These specialized agents could be designed to perform distinct functions, such as generating game rules and action sequences and analyzing the consequences of players' moves. By isolating specific contexts, these specialized agents facilitate interaction among the main agents. They can work together to provide more accurate, efficient, and coherent content generation. For example, by distributing specific parts of the context and isolating the others, a certain agent can specialize in tracking the logical relationship between game states and player actions, while other agents may focus on generating descriptions and reactions that align with the specific game environment or scenario and action sequences of the Gadgets.

% 通过这样的多agent协作系统，我们不仅能够提高模型在复杂场景下的性能，提升速度，以及节省token消耗以降低推理成本。每个agent都可以在其专长领域内提供深入和精准的处理，从而使整个系统能够更好地理解和反映复杂的游戏逻辑和玩家互动。
Through such a two-agent collaboration system, we can not only enhance the agents' performance in complex game scenarios, improve content generation speed, and reduce token consumption to lower inference costs but also allow each agent to provide deep and precise processing within their areas of expertise. This enables the entire SUI system to better understand and reflect the complex logic of games and player interactions.

\subsection{Beyond Action Planning}
This paper primarily focuses on the integration of SUI (Spatial User Interface) with LLM-based (Large Language Model-based) agents in tabletop game scenarios. However, the HCI (Human-Computer Interaction) field has already introduced many instances where SUI is combined with other interactive scenarios, including AR (Augmented Reality) games \cite{sketched-reality}, serious games \cite{pengyu}, and remote interactions \cite{holobots}. This reveals the potential for our AI-Gadget Kit to inspire further research and applications of SUI in scenarios with complex interaction tasks. In this paper, we take the first step by incorporating LLM-based agents for automated action planning in SUI. Future expansions of our kit could enhance understanding and generation of SUI's other interactive modalities. For instance, by integrating Add-ons like the Mechanical Shell \cite{HERMITS}, our kit could transform SUI action planning into planning for other interactivities (e.g., Multi-DoF, Aggregation). Similarly, by generating virtual information, our kit could extend SUI action planning into dynamic interactions that blend virtual and real elements \cite{realitysketch}. Moving forward, we aim to explore more LLM-generated SUI interaction modalities, advancing research and application of SUI in scenarios with complex interaction tasks.

\subsection{Availability and Applicability}
% 我们提出的套件可以帮助相关研究人员、桌游设计师以及玩家以自然语言的方式对自动化的gadget进行创意的交互设计。通过参考套件中two-agent system，以及各类add-on prompt的设计，使用者可以很轻易的对kit中，交互行动序列的生成规则进行调整（例如，在Open-Ai，GPT-4的网页客户端即可进行测试）。
Our proposed AI-Gadget Kit aims to assist researchers, board game designers, and players in creatively designing interactive experiences with automated gadgets using natural language. By leveraging the two-agent system and various add-on prompts included in the kit, users can easily modify the rules for generating sequences of interactive actions (for example, testing can be conducted directly on the OpenAI GPT-4 web client).

% 然而，我们也发现了对Kit生成的行动序列的动态效果进行快速调试时遇到的障碍。
% 在本文中，使用者必须根据不同的场景联通对应数量的移动机器人进行实机测试，或根据行动序列的内容在坐标系上描绘出对应的轨迹，以判断agent生成内容的情况。
% 这阻碍了使用者对创意交互的想法的可视化。
However, we have identified obstacles encountered when attempting to rapidly debug the dynamic effects of action sequences generated by the Kit. Users must either conduct live tests with a corresponding number of gadgets for different scenarios or plot the trajectories on a coordinate system based on the content of the action sequences to evaluate the agent-generated content. This impedes the visualization of creative interactive ideas of the users.

% 最近，WRLKits[WRLKits]演示了一种基于交互式计算设计方法（interactive computational design approach ）的工具，它可视化地帮助了设计人员快速构建个性化原型。
% 同时，Habitat-Sim[24]、AI2-THOR[25]等工作也证明了仿真模拟平台对机器人交互设计的可行性。
% 未来我们将尝试为AI-Gadget Kit开发仿真和设计工具，从而使该套件可以在设计个性化交互时具备在网页或客户端的可视化，进而更容易被许多人访问和使用。
Recently, WRLKits\cite{WRLKits} demonstrated a tool based on the interactive computational design approach, which visually assists designers in rapidly building personalized prototypes. Similarly, works like Habitat-Sim\cite{habitat} and AI2-THOR\cite{AI2-THOR} have proven the viability of simulation platforms for robot interaction design. In the future, we plan to develop simulation and design tools for the AI-Gadget Kit, enabling visualizations on web or client platforms for designing personalized interactions, thus making it more accessible and usable by many.

% 此外，我们也发现目前可以驱动gadget进行交互的SUI的硬件成本，会让这项工作难以广泛的推广到桌游场景中。这是因为大部分桌游店铺很难负担得起高昂的硬件平台（如 Sony Toio平台）费用。
% 为了进一步推广这项工作，未来我们将尝试为AI-Gadget Kit开发更低成本的硬件平台。
In addition, we have noted that the hardware costs associated with SUI capable of driving gadgets for interaction may hinder the widespread adoption of this work in board game scenarios. This is because most board game café may find it challenging to afford the costs of bulk purchasing expensive hardware platforms, such as the Sony Toio platform. To further promote our work, we will explore the development of lower-cost hardware platforms for the AI-Gadget Kit in the future.
\section{Conclusion}
While Swarm User Interfaces (SUIs) have succeeded in enriching tangible interaction experiences, their limitations in autonomous action planning have hindered the potential for personalized and dynamic interaction generation in tabletop games.
In this paper, We proposed an AI-Gadget Kit, a multi-agent SUI tabletop gaming system, which is designed to facilitate dynamic and complex interaction tasks in tabletop games.
We first introduced the system architecture of the AI-gadget Kit, which includes a set of swarm robots to perform the gadget behaviors, and a multi-agent system responsible for executing the game and generating action plans for the swarm robots.
We then elaborated the design of the multi-agent system, comprising a series of meta-motions for individual robots, two LLM-based agents for complex action planning, and a set of add-on prompts aimed at reinforcing the understanding and reacting capabilities of the agents.
At last, we demonstrate four application examples using AI-gadget Kit to showcase the effect of the multi-agent-driven SUI on executing complex interaction tasks in tabletop games.
We aimed to use our work as a case study to explore and inspire the application of LLMs on action planning of SUI in multiple scenarios with complex interaction tasks. 

\bibliographystyle{ACM-Reference-Format}
\bibliography{manuscript}

\appendix


\end{document}